\documentclass[useAMS,usenatbib,usegraphicx]{mn2e}

\usepackage{amsmath, amssymb, graphics, graphicx}
\usepackage[hang,scriptsize]{subfigure}

\newcommand{\mathsym}[1]{{}}

\newcommand{\fref}[1]{Figure~\ref{#1}}
\newcommand{\tref}[1]{Table~\ref{#1}}

\newcommand{\cref}[1]{Chapter~\ref{#1}}

\title[VLBI Polarization of High--z Jets]
{Multi-frequency polarization properties of ten quasars on deca-parsec scales at $z>3$}

\author[O'Sullivan, Gabuzda \& Gurvits]{S. P. O'Sullivan$^{1}$, D. C. Gabuzda$^{2}$ \& L. I. Gurvits$^{3,4}$ \\
$^{1}$CSIRO Astronomy and Space Science, ATNF, PO Box 76, Epping, NSW 1710, Australia.\\
$^{2}$Physics Department, University College Cork, Cork, Ireland.\\
$^{3}$Joint Institute for VLBI in Europe, PO Box 2, 7990 AA Dwingeloo, The Netherlands.\\ 
$^{4}$Institute of Space and Astronautical Science, Japan Aerospace Exploration Agency, 3-1-1 Yoshinodai Chuo-ku, Sagamihara, Kanagawa 252-5210, Japan.\\ }

\begin{document}

\date{Released 2010 Xxxxx XX}
\pagerange{\pageref{firstpage}--\pageref{lastpage}} \pubyear{2010}
\maketitle
\label{firstpage}
\begin{abstract}
Global VLBI (EVN+VLBA) polarization observations at 5 and 8.4 GHz of ten high redshift ($z>3$) 
quasars are presented. The core and jet brightness temperatures are found through modelling the 
self-calibrated \emph{uv}--data with Gaussian components, which provide reliable 
estimates of the flux density and size of individual components. The observed high core brightness 
temperatures (median $T_{\rm b,\,core}=4\times10^{11}$ K) are consistent with Doppler boosted 
emission from a relativistic jet orientated close to the line-of-sight. This can also explain the dramatic 
jet bends observed for some of our sources since small intrinsic bends can be significantly amplified 
due to projection effects in a highly beamed relativistic jet. We also model-fit the polarized emission 
and, by taking the minimum angle separation between the model-fitted polarization angles at 5 and 8.4~GHz, 
we calculate the minimum inferred Faraday rotation measure (RM$_{\rm min}$) for each component. 
We also calculate the minimum intrinsic RM in the rest frame of
the AGN (RM$_{\rm min}^{\rm intr}$ = RM$_{\rm min} (1+z)^2$), first subtracting the integrated (presumed foreground) RM in those cases where we felt we could do this reliably. The resulting
mean core $|$RM$_{\rm min}^{\rm intr}|$ is 5580~rad m$^{-2}$, with a standard deviation of 3390~rad m$^{-2}$, for four high-z quasars for which we believe we could reliably remove the
foreground RM. 
We find relatively steep core and jet spectral 
index values, with a median core spectral index of $-0.3$ and a median jet spectral index of $-1.0$. Comparing our 
results with RM observations of more nearby Active Galactic Nuclei at similar emitted frequencies does not provide any significant evidence 
for dependence of the quasar nuclear environment with redshift. However, more accurate RM and spectral 
information for a larger sample of sources would be required before making any conclusive statements 
about the environment of quasar jets in the early universe. 

\end{abstract}

\section{Introduction} 
Quasars at $z>3$ and with GHz flux densities of the order of 1 Jy have 
luminosities of $\sim10^{28}$~W~Hz$^{-1}$, making them the most luminous, steady radio emitters in 
the Universe. In this paper, we present 5 and 8.4~GHz global VLBI polarization observations of ten 
quasars at $z>3$. There have been several recent studies of high-redshift jets on VLBI scales 
\citep[e.g.,][]{gurvits2000, frey2008, veres2010} but our observations are the first to explore this 
region of ``luminosity -- emitted frequency'' parameter space in detail with polarization sensitivity. 
In the standard $\Lambda$CDM cosmology (H$_0 = 70$ km s$^{-1}$ Mpc$^{-1}$, $\Omega_{\Lambda}=0.7$ 
and $\Omega_{M}=0.3$), the angular diameter distance reaches a maximum at $z\sim1.6$ \citep[e.g.,][]{hogg1999}, 
meaning that more distant objects actually begin to increase in apparent angular size with increasing $z$. 
As a consequence, the linear scale for observations of objects at $z\sim3$ is similar to objects at $z\sim0.7$. 
Therefore, our measurements allow us to study pc-scale structures at emitted frequencies of $\sim$20--45~GHz in 
comparison with properties of low-redshift core-dominated Active Galactic Nuclei (AGN) at 22 and 43~GHz 
\citep[e.g.,][]{osullivangabuzda2009b} at matching length scale and emitted frequency. 

An important question we try to address with these observations is whether or not quasar jets and their 
surrounding environments evolve with cosmic time. Any systematic differences in the polarization properties 
or the frequency dependence of the polarization between high and low redshift sources would suggest that 
conditions in the central engines, jets, and/or surrounding media of quasars have evolved with redshift. 
Previous observations of high redshift quasars \citep[e.g.,][]{frey1997, paragi1999} often show that these 
quasars can be strongly dominated by compact cores and that their extremely high apparent luminosities are 
likely due to large Doppler boosting, implying that their jets move at relativistic velocities and are pointed 
close to our line of sight. With this geometry, it is difficult to distinguish emerging jet components 
from the bright core. However, high resolution polarization observations have proven very effective in 
identifying barely resolved but highly polarized jet components, whose intensity is swamped by the 
intensity of the nearby core, but whose polarized flux is comparable to or even greater than that of 
the core \citep[e.g.,][]{gabuzda1999}. Thus, polarization sensitivity can be crucial to understanding 
the nature of compact jets in the highest redshift quasars.

There is also some evidence that the VLBI jets of high redshift quasars are commonly bent or distorted; one 
remarkable case revealed by VSOP observations is of 1351-018 \citep{gurvits2000} where the jet appears 
to bend through almost $180^{\circ}$. Polarization observations can help elucidate the physical origin of 
these observed changes in the jet direction in the plane of the sky since the direction 
of the jet polarization can reflect the direction of the underlying flow.  In some quasars  
\citep{cawthorne1993, osullivangabuzda2009}, the observed polarization vectors follow curves 
in the jet, indicating that the curves are actual physical bends, while in others, remarkable uniformity 
of polarization position angle has been observed over curved sections of jet, suggesting 
that the apparent curves correspond to bright regions in a much broader underlying flow. 
One possible origin for bends could be collisions of the jet with denser areas in the surrounding 
medium as suggested, for example, in the case of the jet in the radio galaxy 4C~41.~17 \citep{gurvits1997}. 
The polarization information allows us to test for evidence of such interaction, where the 
Faraday rotation measure (RM) and degree of polarization should increase substantially in those regions. 

In Section 2, we describe our observations and the data reduction process. Descriptions of the 
results for each source are given in Section 3, while their implications for the jet structure and environment 
are discussed in Section 4. Our conclusions are presented in Section 5. We use the $S\propto\nu^{\alpha}$ 
definition for the spectral index.

\section{Observations and Data Reduction}
Ten high-redshift ($z>3$) AGN jets (listed in \tref{basic_info}), which were all previously successfully 
imaged using global VLBI baselines, were targeted for global VLBI polarization observations at 4.99 and 8.415~GHz 
with two IFs of 8~MHz bandwidth at each frequency. These sources do not comprise a complete sample in any
sense, and were chosen based on the previous VLBI observations of \cite{gurvits2000}, which indicated the presence of components with sufficiently high total intensities to suggest that it was feasible to detect their
polarization. 
 
The sources were observed for 24 hours on 2001 June 5 with all 10 antennas of the NRAO Very Long Baseline 
Array (VLBA) plus six antennas of the European VLBI Network (EVN) 
capable of fast frequency switching between 5 and 8.4~GHz (see \tref{telescopes} for a list of the 
telescopes used). The target sources were observed at alternating 
several-minute scans at the two frequencies. This led to practically simultaneous observations at the 
two frequencies with a resolution of $\sim1.1$ mas at 5~GHz and $\sim0.7$ mas at 8.4~GHz 
(for the maximum baseline of 11,200 km from MK to NT). 

\begin{table*}
   \caption{Basic information for the high-redshift quasars and VLBI data presented in this paper.}
\smallskip
  \scriptsize{
  \begin{tabular}{ccccccccccccccccccc}
  \hline
 {(1)}                   &  {(2)}                  &
 {(3)}                   &  {(4)}                  &
 {(5)}                   &  {(6)}                  &
 {(7)}                   &  {(8)}      &  {(9)}    &  {(10)}  \\
  \smallskip
   {Name}                  &  {z}         &  {Linear Scale}          &
 {Freq.}         &  {Beam}    & {BPA}   &
 {$I_{\rm total}$}                   &  {$I_{\rm peak}$}                  &
 {$\sigma_{I,\rm{rms}}$}       & $L_{\nu}$ \\  
\smallskip
 { }                      &  { }          &  {[pc mas$^{-1}$]}             &
 {[GHz]}              &  {[mas$\times$mas]}        & {[deg]}          &
 {[mJy]}                  &  {[mJy/bm]}                  &
 {[mJy/bm]}          & [$10^{28}$ W Hz$^{-1}$]\\
\noalign{\smallskip}
  \hline
0014$+$813 & 3.38 & 7.55 &  4.99 &  $1.10\times0.95$ & $-85.9$ & 861.4   &    583.2    &  0.3  &  2.10\\ 
                     &         &         &  8.41 &  $0.95\times0.68$ &  $-85.1$ & 645.1   &   445.6     &  0.9  &  1.57\\ 
                     \noalign{\smallskip}
0636$+$680 & 3.17 & 7.71 &  4.99 &  $1.25\times0.91$ &  $-1.3$ & 348.8   &    270.0    &  0.3  &  0.76\\ 
                     &         &         &  8.41 &  $0.82\times0.62$ &   $5.3$ & 281.4   &    162.7    &  0.5  &  0.62\\ 
                     \noalign{\smallskip}
0642$+$449 & 3.41 & 7.53 &  4.99 &  $1.70\times0.85$ & $-15.6$ &  2248.2   &   1974.9  &  0.7 &  5.55\\
                     &         &         &  8.41 &  $1.03\times0.70$ & $14.4$ &  2743.9   &   2460.4  &  3.9 &  6.77\\
                     \noalign{\smallskip}                      
1351$-$018  & 3.71 & 7.30 &  4.99 &  $3.51\times1.10$ & $-7.0$ &  788.0   &    662.2    &  0.4  & 2.23\\
                     &         &         &  8.41 &  $2.07\times0.86$ & $-2.8$ &  594.0   &    534.2    &  1.4  &  1.68\\ 
                     \noalign{\smallskip}
1402$+$044 & 3.21 & 7.68 &  4.99 &  $3.52\times0.95$ & $-7.7$ &  831.7   &   617.4     &  0.5  &  1.86\\
                     &         &         &  8.41 &  $2.03\times0.75$ & $-6.8$ &  719.7   &    488.2    &  0.3  &  1.61\\
                     \noalign{\smallskip}
1508$+$572 & 4.30 & 6.88 &  4.99 &  $1.23\times0.97$ & $15.2$ &  291.1   &   248.7     &  0.2  &  1.04 \\
                     &         &         &  8.41 &  $0.80\times0.76$ & $8.9$ &  198.3   &    170.3    &  0.5  &  0.71 \\
                     \noalign{\smallskip}
1557$+$032 & 3.90 & 7.16 &  4.99 &  $3.81\times0.97$ & $-8.2$ &  287.0   &    249.6    &  0.3  &  0.88 \\
                     &         &         &  8.41 &  $2.06\times0.80$ & $-8.8$ &  255.8   &    239.1    &  1.0  &  0.78 \\
                     \noalign{\smallskip}
1614$+$051 & 3.21 & 7.68 &  4.99 &  $3.74\times0.98$ & $-8.1$ &  792.9   &    588.9    &  0.5  &  1.77 \\ 
                     &         &         &  8.41 &  $1.99\times0.80$ & $-9.4$ &  465.0   &    333.6    &  1.0  &  1.04 \\ 
                     \noalign{\smallskip}
2048$+$312 & 3.18 & 7.70 &  4.99 &  $2.17\times1.12$ & $-21.2$ &  547.0   &    192.7    &  1.5  &  1.20 \\
                     &         &         &  8.41 &  $2.06\times0.71$ & $-16.2$ &  539.1   &    253.9    &  0.6  &  1.19 \\
                     \noalign{\smallskip}
2215$+$020 & 3.55 & 7.42 &  4.99 &  $3.45\times1.03$ & $-7.6$ &  271.0   &    190.6    &  0.2  &  0.71 \\
                     &         &         &  8.41 &  $2.06\times0.71$ & $-6.8$ &  174.3   &    122.6    &  0.3  &  0.46 \\

\hline
\smallskip
  \end{tabular} \\
  \noindent Column designation: 1~-~source name (IAU B1950.0); 
  2~-~redshift;
  3~-~projected distance, in parsecs, corresponding to 1 mas on the sky; 
  4~-~observing frequency in GHz; 
  5~-~size of convolving beam in mas$\times$mas at the corresponding frequency for each source; 
  6~-~beam position angle (BPA), in degrees;
  7~-~total flux density, in mJy, from \textsc{CLEAN} map;
  8~-~peak brightness, in mJy beam$^{-1}$; 
  9~-~brightness rms noise, in mJy beam$^{-1}$, from off-source region;
  10~-~monochromatic total luminosity at the corresponding frequency in units of $10^{28}$ W Hz$^{-1}$.
  \label{basic_info}
  }
\end{table*}

\begin{table}
\caption{List of telescopes included in the global VLBI experiment.}
\label{telescopes}
\smallskip
\begin{center}
\begin{tabular}{lccl} \hline
  Telescope Location    & Code & Diameter & Comment \\
                    &                     & [m]   &   \\
\hline
\multicolumn{3}{c}{EVN} \\
\hline
Effelsberg, Germany & EB & 100  & \\
Medicina, Italy     & MC & 32   &      \\
Noto, Italy         & NT & 32   &         \\
Toru\'n, Poland       & TR & 32  & Failed     \\
Jodrell Bank, UK     & JB & 25    &  \\
Westerbork, Netherlands & WB & 93$^*$ &\\
\hline
\multicolumn{3}{c}{VLBA, USA}  \\
\hline
Saint Croix, VI         & SC & 25  & \\
Hancock, NH             & HN & 25  &\\
North Liberty, IA       & NL & 25  & \\
Fort Davis, TX          & FD & 25  & \\
Los Alamos, NM          & LA & 25  & Failed \\
Pietown, NM             & PT & 25  & \\
Kitt Peak, AZ           & KP & 25  & \\
Owens Valley, CA        & OV & 25  &\\
Brewster, WA            & BR & 25  &\\
Mauna Kea, HI           & MK & 25  & \\
\hline
\end{tabular}\\
\smallskip
\scriptsize{$^*$ Equivalent diameter for the phased array, $\sqrt{14}\times 25$~m.}
\end{center}
\end{table}

The data were recorded at each telescope with an aggregate 
bit rate of 128 Mbits/s, recorded in 8 base-band channels at 16 Msamples/sec with 2-bit sampling, and 
correlated at the Joint Institute for VLBI in Europe, Dwingeloo, The Netherlands.
Los Alamos (LA) was unable to take data due to communication problems and 
Toru\'n (TR) also failed to take data, so both were excluded from the processing. 

The 5 and 8.4~GHz VLBI data were calibrated independently using standard techniques in the
NRAO AIPS package. In both cases, the reference antenna for the VLBI observations was the Kitt Peak 
telescope. The instrumental polarizations (`D-terms') at each frequency were determined using 
observations of 3C\,84, using the task
\textsc{LPCAL} and assuming 3C\,84 to be unpolarized. The polarization angle ($2\chi=\arctan \frac{U}{Q}$) calibration was achieved 
by comparing the total VLBI-scale polarizations observed for the compact sources 1823+568 (at 5~GHz) 
and 2048+312 (at 8.4~GHz) with their polarizations measured simultaneously with the NRAO 
Very Large Array (VLA) at both 5 and 8.4~GHz, and rotating the VLBI 
polarization angles to agree with the VLA values. We obtained 2 hours of VLA data, overlapping with
the VLBI data on June 6. The VLA angles were calibrated using the known 
polarization angle of 3C\,286 at both 5 and 8.4~GHz (see \textsc{AIPS} 
cookbook\footnote{http://www.aips.nrao.edu/cook.html}, chapter 4). 
At 5~GHz, we found $\Delta\chi_{\rm 5\,GHz}=\chi_{\rm VLA}-\chi_{\rm VLBI}=67.5^{\circ}$ and 
at 8.4~GHz, $\Delta\chi_{\rm 8.4\,GHz}=\chi_{\rm VLA}-\chi_{\rm VLBI}=83.5^{\circ}$. 

\section{Results}
\begin{figure*}
    \includegraphics[width=10cm]{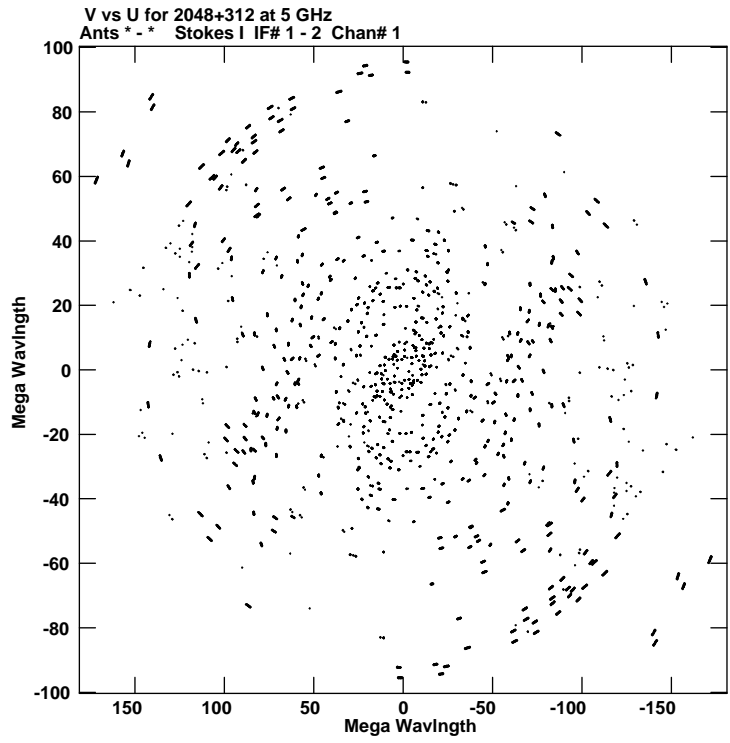}
  \caption[Example plots of the \emph{uv} coverage.]{Example plot of the \emph{uv} coverage obtained for a representative source, 2048+312 at 5~GHz. 
  }
  \label{Figure:uvcoverage}
\end{figure*}

\tref{basic_info} lists the 10 high-z quasars, their redshifts, the FWHM beam sizes at 
5 and 8.4~GHz, the total CLEAN flux and the peak flux densities of the 5 and 8.4~GHz maps, the noise 
levels in the maps, and the 5 and 8.4~GHz luminosities, calculated assuming a spectral 
index of zero and isotropic radiation.  
Since the emission from these sources is likely relativistically boosted, the true luminosity
is probably smaller by a factor of 10--100 \citep[e.g.,][]{cohen2007}. 
\fref{Figure:uvcoverage} gives an example of the sparse but relatively uniform \emph{uv} coverage 
obtained for this experiment. The 5-GHz \emph{uv} coverage for 2048+312 is shown; the 8.4~GHz is 
essentially identical but scaled accordingly. 

The \textsc{DIFMAP} software package \citep{shepherd1997} was used to fit circular and/or elliptical Gaussian 
components to model the self-calibrated source structure. 
The brightness temperatures in the source frame were calculated for each component from the 
integrated flux and angular size according to 

\begin{equation}
T_b=1.22\times 10^{12}\frac{S_{\text{total}}(1+z)}{d_{\max }d_{\min }\nu ^2}
\end{equation}

\noindent where the total flux density is measured in Jy, the FWHM size is measured in mas and the 
observing frequency is measured in GHz. The limiting angular size for a Gaussian component 
($d_{\rm lim}$) was also calculated to check whether the component size reflects the true size of that 
particular jet emission region or not. From \cite{lobanov2005}, we have  

\begin{equation}
d_{\rm lim}=\sqrt{\frac{\ln 2}{\pi} b_{\rm max} b_{\rm min} \ln \left(\frac{\rm S/N}{{\rm S/N}-1} \right)}
\end{equation}

\noindent where $b_{\rm max}$ and  $b_{\rm min}$ are the major and minor axes of the FWHM beam and 
S/N is the signal-to-noise ratio of a particular component. In other words, component sizes smaller
than this value yielded by the model-fitting are not reliable. 
\tref{modelfits} shows the results of the total 
intensity model-fitting for each source. For clarity of comparison, the component identified as the 
core has been defined to be at the origin, and the positions of the jet components determined relative
to this position.  
The errors in the model-fitted positions were estimated as 
$\Delta r = \sigma_{\rm rms}d \sqrt{1+S_{\rm peak}/\sigma_{\rm rms}}/2S_{\rm peak}$ and 
$\Delta\theta = \arctan(\Delta r/r)$, 
where $\sigma_{\rm rms}$ is the residual noise of the map after the subtraction of the model, 
$d$ is the model-fitted component size and $S_{\rm peak}$ is the peak flux density \citep[e.g.,][]{fomalont1999, lee2008, nadia2010}. These
are formal errors, and may yield position errors that are appreciably underestimated in some cases; 
when this occurs we use a $1\sigma$ error estimate of $\pm0.1$ beamwidths in the structural position angle
of the component in question. 

Polarization model fits, listed in Table 4, were found using the 
Brandeis \textsc{VISFIT} package \citep{roberts1987, gabuzda1989} adapted to run in a linux
environment by \citet{bezrukovs2006}. The positions in Table~4 have been shifted in accordance with
our cross-identification of the corresponding intensity cores, when we consider this cross-identification
to be reliable.
The errors quoted for the model-fitted components are formal $1\sigma$ errors, 
corresponding to an increase in the best fitting $\chi^2$ by unity; again, we have adopted a 1$\sigma$
error estimate of $\pm0.1$ beamwidths in the structural position angle
of the component in question when the position errors are clearly underestimated. 

Before constructing spectral-index maps, images with matched resolution must be constructed at 
the two frequencies. Since the intrinsic resolutions of the 8.4 and 5-GHz images were not very different
(less than a factor of two), we achieved this by making a version of the final 8.4-GHz image with the
same cell size, image size and beam as the 5-GHz image. The two images must also be aligned based
on optically thin regions of the structure: the
mapping procedure aligns the partially optically thick cores with the map origin, whereas we
expect shifts between the physical positions of the cores at the two frequencies. When possible,
we aligned the two images by comparing the positions of optically thin jet components at the two frequencies
derived from model-fits to the VLBI data, or using the cross-correlation technique of Croke \&
Gabuzda (2008). After this alignment, we constructed spectral-index maps 
in AIPS using the task COMB.  

In the absence of Faraday rotation, we expect the polarization angles for corresponding regions at
the two frequencies to coincide; in the presence of Faraday rotation of the electric
vector position angle  (EVPA), which occurs when the polarized radiation passes through regions of 
magnetized plasma, the observed polarization
angles will be subject to a rotation by RM$\lambda^2$, where RM is the Faraday rotation measure
and $\lambda$ is the observing wavelength.
We are not able to unambiguously identify the action of Faraday rotation based on observations at
only two frequencies; however, Faraday rotation provides a simple explanation for differences in
the polarization angles observed at the two frequencies. We accordingly calculated tentative Faraday
rotation measures, $\textrm{RM}_{min} = (\chi_{\rm 5~GHz}-\chi_{\rm 8.4~GHz})/(\lambda^{2}_{\rm 5~GHz}-
\lambda^{2}_{\rm 8.4~GHz})$, based on the results of the polarization model fitting,taking the minimum 
difference between the component EVPAs. We also constructed tentative RM maps in AIPS
using the task COMB. In both cases, if Faraday rotation is
operating, the derived RM values essentially represent a lower limit for the true Faraday rotation
(assuming an absence of $n\pi$ rotations in the observed angles).

With our two frequencies, the n$\pi$ ambiguity in the RM corresponds to $n$ times 1350~rad m$^{-2}$.  
Table 4 lists the values of RM$_{\rm min}$ and the corresponding intrinsic polarization angle 
$\chi_0$, obtained by extrapolating to zero wavelength. However, we emphasize that polarization measurements
at three or more wavelengths are required to come to any firm conclusions about the correct RM and $\chi_0$
values for these sources.  

The main source of uncertainty for the RMs (apart from possible $n\pi$ ambiguities) 
comes from the EVPA calibration, which we estimate 
is accurate to within $\pm3^{\circ}$; this corresponds to an RM error of  $\pm23$ rad m$^{-2}$ 
between 5 and 8.4 GHz. We have also attempted to obtain the correct sign and magnitude of the RM in the 
immediate vicinity of the AGNs by subtracting the integrated RMs derived from lower-frequency VLA measurements 
centered on our sources, obtained from the literature. Note that these represent the integrated RMs 
directly along the line of sight toward our sources, rather than along a nearby sight-line; the typical
uncertainties in such measurements when based on observations at several frequencies near 1--2~GHz are typically
no larger than about $\pm 5$~rad/m$^2$. Because of the 
lower resolution of these measurements and the greater prominence of
the jets at lower frequencies, the polarization detected in such observations usually originates fairly far
from the VLBI core, where we expect the overall RM local to the source to be negligible. Therefore, the
lower-frequency integrated RMs usually correspond to the foreground (Galactic) contribution to the overall
RM detected in our VLBI data, and subtracting off this value should typically help isolate the RM occurring
in the immediate vicinity of the AGN. Due to the extremely high redshifts of these sources it can be very 
important to remove the foreground RM before estimating the intrinsic minimum RM in the source rest frame,
RM$_{\rm min}^{\rm intr}=(1+z)^2({\rm RM}_{\rm min}-{\rm RM}_{\rm Gal})$. Table 4 also lists ${\rm RM}_{\rm Gal}$
and RM$_{\rm min}^{\rm intr}$ for sources which have measured integrated RMs in the literature. We take the 
polarization angles measured at the two frequencies to essentially be equal (negligible RM) within the errors 
if they agree to within $6^{\circ}$, and do not attempt to correct such angles for Galactic Faraday rotation,
since there can be some uncertainty about interpreting integrated RMs as purely Galactic foreground RMs, 
and the subtraction of the integrated values from such small nominal VLBI RM values could lead to erroneous
results in some cases.  

For each source, we present the total-intensity distributions at 5 and 8.4~GHz overlaid with EVPA sticks
in Fig.~2 and also plot the positions of the model-fitted total intensity and polarization components. 
The derived spectral-index and RM$_{\rm min}$ distributions generally do not show unexpected features, 
and are not presented here; these can be obtained by contacting S.P. O'Sullivan directly. 

\begin{table*}
  \begin{center}
 \caption{Total intensity model fit parameters of all sources.}
  \smallskip
    \scriptsize{
  \begin{tabular}{ccccccccccccccccccc}
  \hline
{(1)}                   &  {(2)}                  &
 {(3)}                   &  {(4)}                  &
 {(5)}                   &  {(6)}                  &
 {(7)}                   &  {(8)}      &  {(9)}      &  {(10)}         \\
\noalign{\smallskip}
Name                 & Freq.  & Comp.      &
$r$        & $\theta$      &
$I_{\rm tot}$                   & $d_{\rm max}$    & $d_{\rm min}$            &
$d_{\rm lim}$              & $T_{\rm b}$          \\
\noalign{\smallskip}
                      & [GHz]     &               &
[mas]                 & [deg]                  &
[mJy]                   & [mas]    &  [mas]           &
[mas]                   & [$\times10^{10}$ K]          \\     
  \hline
0014$+$813 &4.99& A &  --                           &      --                              &   $617\pm20$ &  0.53  &  0.32   &   0.03    &  $78.2\pm6.4$ \\
                     &           & B &  $0.66\pm0.10$  &   $-175.7\pm1.2$     &   $196\pm11$ &  0.53  &  0.53   &  0.05    &   $15.2\pm2.3$  \\
                     &           & C & $4.93\pm0.10$   &   $-162.8\pm1.1$     &   $13\pm3$     &   0.92  &  0.92   &  0.21    &   $0.32\pm0.21$ \\
                     &           & D & $6.41\pm0.21$   &   $-171.0\pm1.9$     &   $16\pm4$     &   1.75  &  1.75   &  0.19    &   $0.11\pm0.08$ \\
                     &           & E &  $9.51\pm0.16$  &   $-170.9\pm1.0$     &   $17\pm4$     &   1.49 &   1.49  &   0.18   &   $0.16\pm0.11$ \\
                     \noalign{\smallskip}
                     &   8.41 & A &  --                           &  --                             &    $547\pm36$   &   0.46  &   0.13   &  0.04    &   $65.9\pm11.0$  \\
                     &           & B &  $0.74\pm0.07$  &  $-178.3\pm1.7$   &   $85\pm14$     &   0.40  &   0.40   &   0.09    &  $3.9\pm1.7$    \\
                     \noalign{\smallskip}      
0636$+$680 &   4.99 & A &    --                          &   --                                &    $344\pm15$ &   0.69  &    0.35   &   0.03    &  $28.9\pm3.1$ \\
                     &                & B & $2.22\pm0.54$   &    $140.3\pm9.0$    &      $6\pm3$    &   2.76  &   2.76    &   0.23    &  $0.02\pm0.02$ \\
		     \noalign{\smallskip}
                     &   8.41 & A & -  &  -   &     $286\pm21$   &   0.78  &   0.38   &  0.04    &    $7.4\pm1.4$ \\
                     \noalign{\smallskip}
0642$+$449 &   4.99 & A &    --                     &  --                             &   $2209\pm60$ &    0.41  &    0.28   &   0.02    &  $417\pm28$ \\
                     &           & B & $3.22\pm0.45$   &  $92.5\pm7.5$     &    $45\pm14$   &   2.86  &   2.86   &   0.15    &   $0.12\pm0.11$ \\
		     \noalign{\smallskip}
                     &   8.41 & A &  --                            &  --                            &     $2313\pm134$   &   0.16  &   0.16   &  0.03    &    $731\pm103$ \\
                     &           & A1 & $0.20\pm0.09$  &  $158.1\pm0.8$  &     $458\pm60$     &   0.09  &   0.09   &   0.08    &  $439\pm138$ \\
                     \noalign{\smallskip}                     
1351$-$018  &   4.99 & A &  --                       &  --                               &  $681\pm23$     & 0.77 & 0.23 &  0.05 &   $89.5\pm7.5$ \\
                     &           & B & $1.01\pm0.16$   &  $132.7\pm1.0$    &   $79\pm8$     &   0.48 & 0.48 &  0.14  &  $7.9\pm2.0$ \\
		    &           & C & $1.64\pm0.12$   &   $56.6\pm0.5$      &  $14\pm3$      &  0.17  & 0.17 & 0.32  &  --  \\
		    &           & D & $11.85\pm0.59$   & $-12.9\pm2.8$     &  $17\pm5$      &  4.02  & 4.02 & 0.30   &  $0.02\pm0.02$ \\
		   \noalign{\smallskip}
		    &   8.41 & A &  --                          & --                            &    $578\pm37$  &   0.67 & 0.06 &  0.06  &   $125.9\pm20.1$ \\
		    &           & B &  $0.83\pm0.12$ & $138.5\pm9.7$   &    $13\pm6$    &   0.76 & 0.76 &  0.40   &  $0.18\pm0.19$  \\
	 	    &           & C &  $5.96\pm0.63$ & $-2.1\pm5.8$     &    $3\pm3$      &   2.43  & 2.43 &  1.06   & $0.003\pm0.008$ \\
		   \noalign{\smallskip}
1402$+$044 &   4.99 & A &    --                        &   --                               &    $330\pm17$ &    0.23  &    0.23   &   0.05    &  $130.0\pm16.7$ \\
                     &           & B &   $0.76\pm0.25$   &    $-24.5\pm1.3$     &    $374\pm19$ &   0.46  &   0.46   &   0.04    &   $36.9\pm4.6$ \\
                     &           & C &    $3.33\pm0.15$   &    $-44.4\pm4.5$     &    $19\pm5$   &   1.93   &   1.93   &   0.23    &   $0.11\pm0.08$ \\
                     &           & D &    $9.07\pm0.14$   &    $-46.6\pm0.9$     &    $64\pm9$   &   2.13   &   2.13   &   0.11    &   $0.29\pm0.12$ \\
                     &           & E &  $12.64\pm0.50$   &    $-77.7\pm2.3$     &    $44\pm10$   &   4.72  &   4.72   &   0.18    &   $0.04\pm0.03$ \\
                     \noalign{\smallskip}
                     &   8.41 & A &  --                           &  --                            &     $300\pm19$    &   0.14  &   0.14   &  0.07    &   $111.1\pm17.5$ \\
                     &           & B &  $0.83\pm0.16$  &  $-23.7\pm1.0$   &     $336\pm21$     &   0.43  &   0.43   &   0.06    &  $13.4\pm2.1$ \\
                     &           & C &  $3.23\pm0.29$  &  $-43.3\pm5.5$   &       $13\pm5$    &   1.62   &   1.62   &   0.27    &  $0.03\pm0.04$ \\
                     &           & D &  $9.22\pm0.20$  &  $-46.2\pm1.2$   &       $55\pm11$    &   2.07   &   2.07   &   0.15    &  $0.09\pm0.05$ \\
                     &           & E & $12.83\pm0.32$  & $-77.8\pm1.5$    &       $20\pm7$    &  2.09   &  2.09   &   0.18    &  $0.03\pm0.03$ \\
                     \noalign{\smallskip}
1508$+$572 &   4.99 & A & --                             &   --                              &   $276\pm11$   &   0.45   & 0.28 &  0.03   &  $56.1\pm5.3$\\
		     &           & B &  $2.00\pm0.12$    &  $-178.4\pm3.3$    &   $14\pm3$    &    1.37  & 1.37 &  0.13   &  $0.20\pm0.11$ \\
		      \noalign{\smallskip}
                     &   8.41 & A & --                             &   --                                 &    $196\pm14$ &   0.42  &  0.14 &  0.03  & $30.0\pm5.2$  \\
                     &           & B & $1.94\pm0.14$    &   $-172.1\pm3.7$      &       $6\pm3$  &  0.70  & 0.70  &  0.20  &$0.11\pm0.13$   \\
                      \noalign{\smallskip}   
%
%
1557$+$032 &   4.99 & A &    --                    & --                           & $262\pm12$ & 0.70 & 0.26 & 0.06 & $34.0\pm3.9$\\
                     &           & B & $2.64\pm0.17$ & $139.4\pm2.4$ & $14\pm3$   & 1.24 & 1.24 & 0.26 & $0.22\pm0.13$\\
                      \noalign{\smallskip}
                     &   8.41 & A & --                            & --                            & $252\pm15$ & 0.49 & 0.12  & 0.05  & $35.0\pm5.1$\\
                     &           & B & $2.16\pm0.37$   & $148.5\pm9.2$  & $15\pm5$   & 2.46 & 2.46  & 0.18 & $0.02\pm0.02$  \\
                     \noalign{\smallskip}
1614$+$051 &   4.99 & A &    --                      &   --                                 & $566\pm24$    &    0.37  &    0.37   &   0.05    &  $63.5\pm6.7$ \\
                     &           & B & $1.26\pm0.16$   &    $-154.4\pm0.7$     &  $229\pm15$   &   0.54  &   0.54   &   0.08    &   $12.4\pm2.1$ \\
		     \noalign{\smallskip}
                     &   8.41 & A &  --                           &  --                              &     $347\pm28$     &   0.24  &   0.24   &  0.07    &    $32.4\pm6.5$ \\
                     &           & B &  $1.26\pm0.12$  &  $-154.1\pm0.7$   &     $117\pm17$     &   0.37  &   0.37   &   0.12    &   $4.6\pm1.6$ \\
                     \noalign{\smallskip}   
2048$+$312 &   4.99 & A & --                          & --                         & $393\pm38$ & 1.78 & 1.24 & 0.10 & $3.7\pm1.0$ \\
		     &           & B & $1.55\pm0.11$  & $65.7\pm2.2$ & $126\pm21$ & 1.43 & 1.43 & 0.17 & $1.3\pm0.6$ \\
		     &           & C & $8.86\pm0.36$ &$56.8\pm2.0$ & $20\pm10$   & 1.68 & 1.68 & 0.44 & $0.14\pm0.19$ \\
		      \noalign{\smallskip}
                     &   8.41 & A & --                          & --                              & $411\pm23$ & 0.80 & 0.44 & 0.04  &  $8.4\pm1.2$ \\
                     &           & B & $1.49\pm0.07$ &   $80.6\pm2.6$    & $129\pm14$ & 1.29 & 1.29  & 0.07 &   $0.56\pm0.18$  \\    
                      \noalign{\smallskip}                
2215$+$020 &   4.99 & A &   --                      &  --                              & $141\pm9$ & 0.26  &  0.26   & 0.08   &  $45.9\pm7.3$ \\
                     &           & B & $0.52\pm0.11$   &   $108.0\pm2.2$     &  $73\pm7$  &  0.39  & 0.39   &  0.12 &   $10.7\pm2.4$ \\
                     &           & C &$1.53\pm0.10$    &  $79.1\pm5.4$     &  $11\pm3$  &  1.26  &  1.26   & 0.30  &   $0.15\pm0.10$ \\
                     &           & D &$50.52\pm1.52$  &  $79.1\pm1.6$     &  $33\pm11$  &  9.17  &  9.17   & 0.17  &   $0.009\pm0.009$ \\
                     &           & E & $57.71\pm2.48$  &  $74.7\pm2.3$    &   $19\pm8$ &   11.26 &11.26  & 0.23  &  $0.003\pm0.004$ \\
                     \noalign{\smallskip}
                     &   8.41 & A & --                           &  --                              &    $94\pm7$   &  0.20  & 0.20   &  0.06   &  $19.0\pm3.5$  \\
                     &           & B & $0.50\pm0.08$  &  $115.7\pm1.3$     &  $66\pm6$  &   0.22 &  0.22   &  0.07   &  $10.7\pm2.4$ \\
                     &           & C & $0.95\pm0.07$  &  $95.5\pm3.6$    &     $17\pm3$  &   0.64  & 0.64   &  0.14    & $0.32\pm0.2$  \\
                     \noalign{\smallskip}      
\hline
\smallskip
  \end{tabular}\\
  \noindent Column designation: 1~-~source name (IAU B1950.0); 
2~-~observing frequency, in GHz;
3~-~component identification; 
4~-~distance of component from core, in mas; 
5~-~position angle of component with respect to the core, in degrees; 
6~-~total flux of model component, in mJy;
7~-~FWHM major axis of Gaussian component, in mas; 
8~-~FWHM minor axis of Gaussian component, in mas; 
9~-~Minimum resolvable size, in mas;
10~-~measured brightness temperature in units of $10^{10}$K.
  \label{modelfits}
  }
  \end{center}
\end{table*}


\begin{table*}
  \begin{center}
  \scriptsize{
  \begin{center}
 \caption{Polarization model fit parameters of all sources.}
  \end{center}
  \begin{tabular}{ccccccccccccccccccc}
  \hline
{(1)}                   &  {(2)}                  &
 {(3)}                   &  {(4)}                  &
 {(5)}                   &  {(6)}                  &
 {(7)}                   &  {(8)}      &  {(9)}     &  (10)  &  (11)   & (12) \\
\noalign{\smallskip}
Name                 & Freq.  & Comp.      &
$r$        & $\theta$      &
$p$                   & $\chi$    & $m$ & RM$_{\rm min}$            &
$\chi_0$     &   RM$_{\rm Gal}$   &    RM$_{\rm min}^{\rm intr}$  \\
\noalign{\smallskip}
                      & [GHz]     &               &
[mas]                 & [deg]                  &
[mJy]                   & [deg]  &  [\%]    &  [rad m$^{-2}$]           &
[deg]            &  [rad m$^{-2}$]     &  [rad m$^{-2}$]    \\     
  \hline
0014$+$813 &   4.99 & A & --                           &  --                                &   $4.5\pm0.3$  &  $-40.9$  &  $1.1\pm0.3$    &  384  &  $-120\pm13$  & $9^a$ &  $7194\pm525$  \\
                        &            & B &  $4.72\pm0.10$  &  $-163.5\pm2.1$      &   $1.3\pm0.5$  &   47.7       &  $20.0\pm4.5$ &    --    &    --           & --          &   --   \\
                     \noalign{\smallskip}
                     &   8.41 & A &  --     & --      &    $2.3\pm0.2$   &   $-92.2$  &   $0.7\pm0.2$ &           &                  &             &       \\
                     \noalign{\smallskip}      
0636$+$680 &   4.99 & A &  --   & --   &    $4.1\pm0.3$   &   78.4        &   $1.7\pm0.5$  &    --    &   --            & --          &   --    \\
		     \noalign{\smallskip}
                     &   8.41 & -- &  --  &  --   &     --   &   --  &  -- &   &    &  &    \\
                     \noalign{\smallskip}
0642$+$449 &   4.99 & A &  --   &   --    &   $21.2\pm1.8$ &   15.6        &  $1.1\pm0.4$   &    --    &   --        &  -- &  --   \\
		     \noalign{\smallskip}
                     &   8.41 & A &  --     & --         &     $68.2\pm4.2$ &   15.3       &  $2.6\pm0.7$   &           &               &                 &        \\
                     \noalign{\smallskip}                     
1351$-$018  &   4.99 & A & --                          &  --                              &  $5.2\pm0.4$       &  $-76.2$ &   $0.9\pm0.3$  &$-223$ & $ -30\pm6$ & $-8^a$   & $-4770\pm599$  \\
                         &           & B & $1.45\pm0.11$   &  $100.3\pm2.7$       &   $2.8\pm0.7$      &   52.3     &    $2.7\pm0.5$   &   199    &   $11\pm3$       & $-8^a$  &  $4592\pm646$ \\
		   \noalign{\smallskip}
		    &   8.41 & A & --                         &  --                                  &    $4.1\pm0.4$     &   $-46.4$ &   $0.5\pm0.2$   &              &               &                &    \\
		    &           & B &  $2.38\pm0.16$ &   $155.9\pm2.4$         &    $1.7\pm0.5$     &   25.7       &  $1.3\pm0.4$    &              &                &                &    \\
		   \noalign{\smallskip}
1402$+$044 &   4.99 & A &  --                           &   --                       &   $3.1\pm0.3$      &  15.7        &  $1.8\pm0.5$    &  $-199$ &    $57\pm12$     &        --     &   --  \\
                     &               & B &  $0.77\pm0.11$   & $-61.5\pm1.8$    &  $11.2\pm1.1$     & 4.1           &  $2.0\pm0.7$    &   $-57$   &    $16\pm12$     &     --        &   --  \\
                     &               & C & $1.50\pm0.12$   & $-60.9\pm3.3$  &  $1.6\pm0.5$       &  $-26.2$  &  $2.9\pm0.8$     &    141    & $-55\pm15$    &      --      &   --    \\
                     &               & D &  $8.12\pm0.15$  & $-45.4\pm0.9$  &  $1.1\pm0.4$       &  83.8        &  $17.8\pm5.4$  &  $-59$   &   $96\pm49$        &      --      &   --    \\
                     &               & E &  $9.56\pm0.14$   & $-49.2\pm1.2$  &  $2.0\pm0.5$      &  $-1.5$    &  $9.1\pm2.1$     &    --         &       --      &      --      &   --       \\
                     &               & F &  $13.41\pm0.10$ & $-81.9\pm1.7$  &  $1.6\pm0.5$      & 49.8         &  $21.2\pm6.2$  &    --         &        --     &     --       &   --        \\
                     \noalign{\smallskip}
                     &   8.41 & A1 &  $2.77\pm0.20$  & $177.5\pm3.4$ &  $5.8\pm0.6$      &  $-6.2$     &  $3.2\pm0.7$  &                &                &               &    \\
                     &           & A    & --                           & --                          &   $2.3\pm0.5$     &   42.3        &   $1.5\pm0.3$ &               &                &               &    \\
                     &           & B   & $0.54\pm0.09$   & $-60.1\pm2.7$     &    $7.4\pm0.7$     &  11.7        &    $1.3\pm0.3$ &               &                &                &   \\
                     &           & C   &  $1.48\pm0.08$ & $-72.3\pm2.4$   &     $2.8\pm0.6$    &  $-45.1$  &    $1.2\pm0.3$  &               &                  &              &   \\
                     &           & D   & $8.26\pm0.11$  & $-43.0\pm0.9$  &      $2.8\pm0.6$    &  91.7        &   $22.9\pm9.1$ &             &                   &               &        \\
                     \noalign{\smallskip}              
1508$+$572 &   4.99 & A & --                            &  --                               &   $7.5\pm0.9$   &   $-89.4$   & $2.9\pm0.8$  &    --    &  --   &  -- & --  \\
		     &           & B &  $2.36\pm0.18$    &  $177.3\pm1.4$       &   $0.6\pm0.3$    &    36.1       &  $16.9\pm5.7$ &    --   &    --         &     --         &   --     \\
		      \noalign{\smallskip}
                     &   8.41 & A &  --                            &   --                                &    $4.0\pm1.2$  &   $-93.4$  &  $1.3\pm0.4$   &          &               &                  &       \\
                      \noalign{\smallskip} 
1557$+$032 &   4.99 & A & --                          & --                                 &  $3.4\pm0.5$     & 54.2          &  $1.7\pm0.5$   & $-49$  &  $64\pm40$    & 3$^a$     &  $-1249\pm713$ \\
                         &           & B &  $1.77\pm0.19$ &   $19.5\pm3.6$         & $0.7\pm0.3$      & 85.2          &  $1.8\pm0.5$   &    68     &  $71\pm32$     & 3$^a$     &   $1561\pm642$ \\
                      \noalign{\smallskip}
                     &   8.41 & A & --                           & --                                 & $3.5\pm0.4$         & 60.8         &   $1.2\pm0.4$  &              &              &                &   \\
                     &           & B &  $1.25\pm0.09$   & $59.1\pm1.7$          & $2.1\pm0.3$         & 76.2         &  $6.0\pm0.7$   &              &               &               &    \\
                      \noalign{\smallskip}
1614$+$051 &   4.99 & C &  $1.51\pm0.10$ & $-95.7\pm1.0$  & $3.8\pm0.5$      &  $-38.5$  &  $1.8\pm0.5$   &   522    &  $-107\pm10$   & 8$^c$ &  $9110\pm488$  \\                     
		     \noalign{\smallskip}
                     &   8.41 & A &  --                         &  --                            &     $2.9\pm0.6$        &   62.4       &   $3.4\pm0.7$  &              &                  &                 &     \\
                     &           & B &  $0.82\pm0.08$  & $-119.1\pm2.9$   &    $6.5\pm0.9$     &  78.7        &   $2.1\pm0.4$  &             &                    &                   &    \\
                     &           & C &  $1.55\pm0.09$ &  $-121.9\pm3.2$   &     $2.6\pm0.5$     & $-69.8$  &   $2.0\pm0.3$  &            &                     &                  &    \\
                      \noalign{\smallskip}   
2048$+$312 &   4.99 & A & --                         & --                          & $3.8\pm0.6$            & $-39.5$  &  $3.3\pm0.2$   & -- & --     & -- & --  \\
		     &           & B &  $2.31\pm0.11$ & $88.1\pm1.6$     & $1.1\pm0.5$            &  $-61.2$ &  $4.7\pm0.4$   & -- & -- & -- &   --    \\
		      \noalign{\smallskip}
                     &   8.41 & A & --                        & --                          & $5.5\pm0.7$           & $-35.0$   &   $2.2\pm0.3$    &           &                  &                     &   \\
                     &           & B & $2.04\pm0.09$ &  $86.6\pm1.9$  & $1.4\pm0.5$           & $-93.6$    &   $6.2\pm0.6$   &          &                   &                     &    \\    
                      \noalign{\smallskip}                
2215$+$020 &   4.99 & A & --    &  -- & $0.6\pm0.2$          & 57.0       &   $0.9\pm0.4$     &   --   &       --       & --     &  --   \\
                     \noalign{\smallskip}
                     &   8.41 & A & --   &  --    &  $1.5\pm0.4$      &  53.8        &  $1.3\pm0.5$    &          &                   &                     &   \\
                     \noalign{\smallskip}       
\hline
  \end{tabular}\\
Column designation: 1~-~source name (IAU B1950.0); 
2~-~observing frequency, in GHz;
3~-~component identification; 
4~-~distance of component from core component, in mas; 
5~-~position angle of component with respect to core component, in degrees;
6~-~polarized flux of component, in mJy;
7~-~EVPA of component, in degrees (nominal error of $\pm3^{\circ}$ from calibration, which is much greater than any model-fit errors); 
8~-~Degree of polarization, in per cent, taken from a 3x3 pixel area in degree of polarization image centred on the model-fitted position;
9~-~minimum rotation measure obtained from minimum separation between 
$\chi_{\rm 5~GHz}$ and $\chi_{\rm 8.4~GHz}$, in rad m$^{-2}$ (with nominal error of  $\pm28$ rad m$^{-2}$); 
10~-~intrinsic EVPA, in degrees, as corrected by RM$_{\rm min}$.
11~-~integrated (Galactic) RM, in rad m$^{-2}$, from literature 
$^a$\citet{taylor2009}, $^b$\citet{orenwolfe1995}.
12~-~intrinsic minimum RM corrected for Galactic contribution where 
RM$_{\rm min}^{\rm intr}=({\rm RM}_{\rm min}-{\rm RM}_{\rm Gal})(1+z)^2$.
  \label{pol_models}
  }
  \end{center}
\end{table*}


\begin{table}
\scriptsize{
\caption{Total Intensity and Polarization image cutoff values.}
\label{Table1}
\begin{center}
\begin{tabular}{ccccccccccccc} \hline
\multicolumn{0}{}{}        & \multicolumn{2}{c}{5 GHz} & \multicolumn{2}{c}{8.4 GHz}   \\ \cline{2-3} \cline{4-5} 
Source name   & $I$ cutoff  &  $p$ cutoff   & $I$ cutoff  &  $p$ cutoff\\
                           & [mJy/bm]   &[mJy/bm]      & [mJy/bm]   &[mJy/bm] \\
\hline

0014$+$813  &    1.2      &       0.7              & 2.8              &  1.3   \\ 
0636$+$680  &    0.9      &       0.7              & 1.5              & 0.2   \\
0642$+$449  &    3.6      &       2.4              & 5.0              &  11.0   \\ 
1351$-$018   &    1.3      &       1.1              & 4.0              & 1.3   \\
1402$+$044  &    2.0      &       1.4              & 1.1              &  1.7   \\ 
1508$+$572  &    1.1      &       0.7              & 1.5              & 1.8   \\
1557$+$032  &    1.1      &       0.9              & 1.0              &  1.2   \\ 
1614$+$051  &    1.7      &       1.1              & 4.2              & 1.5   \\
2048$+$312  &    6.0      &       1.9              & 1.9              &  1.5   \\ 
2215$+$020  &    0.7      &       0.7              & 1.6              & 0.9   \\
\hline
\end{tabular}\\
\smallskip
\end{center}
}
\end{table}

\subsection{0014+813}
This source was discovered in the radio and classified as a flat spectrum radio quasar by \citet{kuhr1981}. 
\citet{kuhr1983} obtained a redshift of 3.366, and found the source to be exceptionally bright in the 
optical but unpolarized, while \citet{kaspi2007} found very little optical variability for the source. 
From VLBA observations at 8~GHz over 5 years, \citet{piner2007} did not detect any outward motion. 
The source was observed by the \emph{Swift} satellite in 2007 January from optical to hard 
X-rays. Through construction of the spectral energy distribution (SED), \cite{ghiselliniBHmass2009} show 
that it may harbour one of the largest black hole masses ever inferred. By associating the strong 
optical--UV emission with a thermal origin from a standard optically-thick geometrically-thin accretion 
disk they estimate a black hole mass of $\sim4\times10^{10}M_{\odot}$.

Our 5~GHz image 
shows a core-jet structure extending Southwards to $\sim11$ mas, consistent with the VSOP Space-VLBI image 
at 1.6~GHz of \citet{hirabayashi1998}, which has about a factor of two worse resolution than our own
image in the  North-South direction.
At 8.4~GHz 
the extended jet emission is very faint. If we cross-identify an intensity component in the inner jet with the 
innermost jet component at 5~GHz in order to align our images, we obtain a physically reasonable 
spectral index map. After applying this same relative shift to the polarization model fits, both fits indicate
a polarized component close to the origin, with a residual offset between frequencies of less than 
0.20~mas; we accordingly have identified both components with the core polarization. 
A comparison of the corresponding model-fitted core EVPAs indicates a minimum 
RM of 384 rad m$^{-2}$.

\subsection{0636+680}
This source has been observed previously on mas scales in the radio at 5~GHz by \citet{gurvits1994} 
and more recently by \citet{fey1997} with the VLBA at 2.3 and 8.6~GHz who found it to be unresolved.
Their resolution at 8.6~GHz ($1.3\times1.2$ mas) is similar to our 
resolution at 5~GHz ($1.25\times0.91$ mas). It was first reported as a GPS source by \cite{odea1990}. 
The redshift of 3.17 is quoted from the catalogue of \citet{veron-cettyveron1989}. 

Due to the lack of extended jet emission at 8.4~GHz the images could not be aligned 
based on their optically thin jet emission. 
Polarized emission was detected in the core in our 5~GHz map at the level of $m\sim$1--2\%; no 
polarized emission was detected at 8.4~GHz above 0.2~mJy.  

\subsection{0642+449}
This extremely luminous quasar ($L_{\rm 8.4\,GHz}=6.8\times10^{28}$ W Hz$^{-1}$; the highest in 
our sample) is regularly observed by the MOJAVE\footnote{http://www.physics.purdue.edu/astro/MOJAVE/index.html} 
team with the VLBA at 15~GHz, who have reported subluminal speeds of $\beta\sim0.76$ for an inner 
jet component \citep{lister2009}. A 5-GHz global VLBI image by \citet{gurvits1992} shows the 
jet extending almost 10 mas to the East while \cite{volvach2003} find the jet extending $\sim5$ mas to 
the East from EVN observations at 1.6~GHz. 

Our 5-GHz image is consistent with the images of
 \cite{gurvits1992} and \cite{volvach2000}, with no extended jet emission 
detected at 8.4~GHz.  Due to the lack of extended jet emission at 8.4~GHz the images could not be aligned 
based on their optically thin jet emission. 
Polarized emission is detected in the core at both 5 and 8.4~GHz; the minimum difference between
the model-fitted polarization angles is less than $1^{\circ}$, nominally indicating negligible Faraday
rotation. 

\subsection{1351-018}
This radio loud quasar is the third most distant source in our sample with a redshift of 3.707 
\citep{osmer1994}. It was observed by the VSOP Space-VLBI project at 1.6 and 5 GHz \citep{frey2002}. 
It has a complex pc-scale structure with the jet appearing to bend 
through $\sim130^{\circ}$. The high resolution space-VLBI images ($1.56\times0.59$ mas at 5~GHz) 
clearly resolve an inner jet component within 1 mas of the core with a position angle of 
$\sim 120^{\circ}$, which was also detected at 8.6~GHz by \citet{fey2000}.  

Our 5-GHz image is dominated by the core emission, but the polarization gives a clear indication 
of the presence of an inner jet component, which is also indicated by the intensity model fitting. 
Although two polarized components are visible in the 8.4-GHz map, the relationship between these and
the total-intensity structure is not entirely clear. The polarization angle of the polarized feature
slightly north of the core is similar to the EVPA of the 5-GHz core, and we have on this basis 
identified this polarization with the 8.4-GHz core. The beam is relatively large in the North--South
direction, and it is possible that this has contributed to the shift of this polarization from its
true position relative to the intensity structure. If this identification is correct, the minimum RM 
obtained for the core is $-223$ rad m$^{-2}$. 

\subsection{1402+044}
This flat-spectrum radio quasar has a redshift of 3.208 \citep{barthel1990}. Although the source
structure is fairly complex, the intensity structures match well at the two frequencies. The
components in the innermost jet lie along a position angle of $-24^{\circ}$, consistent with the 
higher resolution image of \cite{yang2008}, who find an inner jet direction of $-26^{\circ}$ from 
5-GHz VSOP data. 

Polarized components A, B, C, and D agree well between the two frequencies, although it is not
clear how these correspond in detail to the intensity components in the same regions. There is
an additional polarized component A1 in the 8.4-GHz core, which does not have a counterpart in
intensity or at 5~GHz; it is difficult to be sure whether this is a real feature or an artefact.

The core polarization structure at both 5 and 8.4 GHz shows three distinct features within the 
core region. If we compare this with the VSOP space-VLBI image of \cite{yang2008} at 5 GHz we see 
that a similar type of structure is seen in total intensity. This indicates how the polarized 
emission can give information about the jet 
structure on scales smaller than is seen in the $I$ image alone. 


\subsection{1508+572}
This is the most distant object in our sample at a redshift of 4.30 \citep{hook1995}. Hence, the 
frequencies of the emitted synchrotron radiation are the highest in our sample at 24.6 and 44.6~GHz. 
The images presented here are the first observations to show the direction of the inner 
pc-scale jet. This quasar also has an X-ray 
jet extending in a south-westerley direction on kpc (arcsecond) scales detected with 
\emph{Chandra} \citep{siemiginowska2003, yuan2003}. \cite{cheung2004} detected a radio jet 
in the same region using the VLA at 1.4~GHz. 

Although an optically thin jet component is detected at both frequencies roughly 2~mas to the
south of the core, this component proved too weak to be used to align the two images. Both the core
and this jet component were detected in polarization at 5~GHz, while only the core was detected 
at 8.4~GHz. The smallest angle between the 5 and 8.4~GHz core EVPAs is only $1^{\circ}$, which we take
to indicate an absence of appreciable Faraday rotation.  

\subsection{1557+032}
This quasar is the second most distant object in our sample with a redshift of 3.90 \citep{mcmahon1994}. 

There are two distinct polarized components in the core region where only one total intensity 
component is distinguished; the fact that these two components are visible and model fit at
both frequencies with similar positions and polarization angles suggests that they are real. 
This appears to be a case when the polarized emission provides information on scales smaller 
than those evident in the total intensity image. 

\subsection{1614+051}
This quasar, at a redshift of 3.212 \citep{barthel1990}, has been observed by \citet{fey2000} with the 
VLBA at 2.3 and 8.6~GHz, who observed the jet to lie
in position angle $-158^{\circ}$. Our 8.4~GHz image 
shows the source clearly resolved into a core and inner jet, extending in roughly this same position angle.
 
Both our 5 and 8.4-GHz data are best fit with two components, whose positions agree well at the two frequencies. One polarized component was model-fit at 5~GHz (Table 4); its position does not completely agree with
the positions of either of the two $I$ components, although it clearly corresponds to jet emission.
Polarization from both of the $I$ components is detected at 8.4~GHz (Table 4), as well as another
region of polarized emission between them. Component C in the 8.4-GHz polarization fit is weak, but
its position agrees with that of the jet $I$ component fit at this frequency, suggesting it may be
real. 

\subsection{2048+312}
\citet{veron-cettyveron1993} found a redshift of 3.185 for this quasar. The apparent shift in the
position of the peak between 5 and 8.4~GHz is an artefact of the mapping process; the model fits indicate
a core and inner jet whose positions agree well at the two frequencies, as well as another weaker jet
component further from the core detected at 5 GHz. 

This is a promising candidate source for follow-up multi-frequency polarization observations because 
the jet is well resolved with VLBI and has a strongly polarized core and jet. 

\subsection{2215+020}
\citet{drinkwater1997} found an emission redshift of 3.572 for this quasar.
\citet{lobanov2001} present a 1.6~GHz VSOP space-VLBI
image of this source showing the jet extending to almost 80 mas ($\sim600$ pc) with a particularly 
bright section between 45 and 60 mas. The extent of this jet is $\sim4$ times greater than in any 
other pc-scale jet observed for quasars with $z>3$. 

Our 5~GHz image 
has similar resolution to the 
\citet{lobanov2001} image but is less sensitive to the extended jet emission hence, our image has a 
similar but sparser intensity distribution. We also detect the particularly bright region, where the jet 
changes direction from East to North-East on the plane of the sky.  

\section{Discussion}
\subsection{Brightness Temperature}
The median core brightness temperature at 5~GHz is $5.6\times10^{11}$ K, while at 8.4~GHz, 
the median value is slightly smaller at $3.2\times10^{11}$ K. This can also be seen from a 
histogram of the core brightness temperature (\fref{fig:Tb_histogram}) where the 8.4~GHz values populate 
a larger majority of the lower bins. 
This may be as a result of the 8.4~GHz data probing regions of the jet where the physical conditions 
are intrinsically different, leading to lower observed brightness temperatures, similar to the results of \cite{lee2008} at 86~GHz.  
However, it is difficult to separate this effect from any bias due to the resolution difference between 5 and 8.4~GHz. 
Furthermore, due to the relatively small difference between the median values, a larger sample size would be required 
to determine whether this is a real effect or just scatter in the data for the small number of sources observed here. 

The maximum core brightness temperature at 5~GHz is $4.2\times10^{12}$ K found in 
0642+449 and the minimum value is $3.7\times10^{10}$ K found in 2048+312. 
At 8.4~GHz, the maximum and minimum values are $7.3\times10^{12}$ K (0642+449) 
and $7.4\times10^{10}$ K (0636+680). Assuming that the intrinsic brightness temperature 
does not exceed the equipartition upper limit of $T_{eq}\sim10^{11}$ K \citep{readhead1994}, 
we can consider the observed core brightness temperatures in excess of this value to be 
the result of Doppler boosting of the approaching jet emission. Using the equipartition 
jet model of \cite{blandfordkonigl1979} for the unresolved core region, we can estimate 
the Doppler factor ($\delta$) required to match the observed brightness temperatures 
($T_{\rm b, obs}\sim3\times10^{11}\delta^{5/6}$). In most cases the Doppler factors 
required are modest, with values ranging from 1 to 5, except for 0642+449 which 
requires a Doppler factor of 23 at 5~GHz and 46 at 8.4~GHz.

From the observed brightness temperatures of the individual jets components, we can investigate the 
assumption of adiabatic expansion following \citet{marscher1990}, \citet{lobanov2001} and 
\citet{pushkarev2008}, who model the individual jet 
components as independent relativistic shocks with adiabatic energy losses dominating the 
radio emission. Note that this description of the outer jet differs from the $N\propto r^{-2}$, $B\propto r^{-1}$ 
case describing the compact inner jet region, which is not adiabatic.  With a power-law energy 
distribution of $N(E)\propto E^{2\alpha-1}$ and a magnetic field described by $B\propto d^{-a}$, where 
$d$ is the transverse size of the jet and $a=1$ or $2$ corresponds to a transverse or longitudinal magnetic field orientation, 
we obtain 

\begin{equation}
T_{\rm b,\,jet}=T_{\rm b,\,core}\left(\frac{d_{\rm core}}{d_{\rm jet}}\right)^{a+1-\alpha(a+4/3)}
\end{equation}

\noindent which holds for a constant or weakly varying Doppler factor along the jet. Hence, we can compare 
the expected jet brightness temperature ($T_{\rm b,\,jet}$) with our observed values by using the 
observed core brightness temperature ($T_{\rm b,\,core}$) along with the measured size of the 
core and jet components ($d_{\rm core}$ and $d_{\rm jet}$) for a particular jet spectral index. 
We apply this model to the sources with extended jet structures to see if this model is an accurate 
approximation of the jet emission and also as a diagnostic of regions where the physical properties 
along the jet may change.  

\fref{Figure:0014_Tb} shows the comparison of this model (solid line) with the observed brightness 
temperatures (dashed line) along the jet of 0014+813 at 5~GHz. From the spectral index 
distribution obtained between 5 and 8.4~GHz, we adopt a jet spectral index of $\alpha=-1.0$ and 
assume a transverse magnetic field orientation ($a=1$) from inspection of the jet EVPAs (Fig.~2).
The two jet components furthest from the core agree relatively well with the model, however, the two 
jet components closest to the core are substantially weaker than expected. Given the high RM for this 
source, it is also possible that the intrinsic magnetic field orientation is longitudinal; using a value of $a=2$ 
yields model brightness temperatures for these components that are within our measurement errors. 
Unfortunately, the extended jet emission at 8.4~GHz is not detected so we cannot constrain the jet RM
and we also cannot rule out a change in the Doppler factor along the jet due to either an intrinsic change in 
the jet dynamics or a change in the jet direction. Clearly there are not enough observational constraints 
for this source to convincingly test the applicability of this model.

The observed brightness temperatures in the jet of 1402+044 at both 5 and 
8.4~GHz (\fref{Figure:1402_Tb}) are in reasonable agreement with the model 
predictions for $\alpha=-0.5$, which is consistent with the overall jet spectral index 
distribution. 
Since the EVPA orientation 
implies a transverse magnetic field (\fref{Figure:I+p}), we use $B\propto d^{-1}$. 
The jet components at 1~mas and 9~mas from the core both have observed brightness temperatures  
slightly higher than would be expected for radio emission dominated by adiabatic losses. Using 
flatter spectral index values of $\alpha=0.0$ and $\alpha=-0.3$, respectively, for these two components
brings the model brightness temperatures within the measurement errors. 
Hence, it is likely that these components are subject to strong reacceleration mechanism that temporarily overcomes the energy losses due to the expansion of the jet.

Applying the analysis to the jet of 2215+020 at 5~GHz (\fref{Figure:2215_Tb}), we see that 
the model matches the observed values very well for the two inner jet components using our 
measured spectral index of $\alpha=-0.7$ 
and a transverse magnetic field ($a=1$). However, the extended components brightness 
temperatures at 50--60~mas from the core are higher than expected by the adiabatic expansion model. 
This suggests different physical conditions in this region of the jet creating a less steep spectral index 
(using $\alpha=-0.5$ brings the model in line with the observed values). 
This far from the core it is likely that the Doppler factor has changed with either a change in the jet 
viewing angle or speed, possibly from interaction with the external medium. This is consistent 
with the observed strong brightening of the jet along with its change in direction from East to 
North-East, shown in \fref{Figure:I+p}. Our results are very similar to those of \citet{lobanov2001} 
who employed the same model for this source at 1.6~GHz. 

\subsection{Jet Structure and Environment}
The added value of polarization observations in providing information on the compact 
inner jet structure is clear from the images of 1402+044 (\fref{Figure:I+p}), 1557+032 (\fref{Figure:I+p}) 
and 1614+051 (\fref{Figure:I+p}), where polarized components are resolved on scales smaller than obtained 
from the total intensity image alone. For 1402+044, a higher resolution VSOP image at 5~GHz 
\citep{yang2008} resolved three components in the compact inner jet region consistent with what we find from the 
polarization structure of our lower resolution images where only one total intensity component is visible.  
In the case of 1557+032 and 1614+051, the identification of multiple polarized components 
in the core region, where again only one total intensity component is directly observed, suggests that 
observations at higher frequencies and/or longer baselines provided by, for example, the 
VSOP-2 mission \citep{tsuboi2009} are likely to resolve the total intensity structure of the core region. 
Hence, the polarization information is crucial in identifying the true radio core and helping to determine whether it 
corresponds to the $\tau=1$ frequency-dependent surface first proposed by \citet{blandfordkonigl1979} 
or a stationary feature such as a conical shock \citep{cawthorne2006, marschervsop2009} for a particular source. 

For the sources with extended jets, substantial jet bends are observed. This is not a surprise since 
as long as the observations imply relativistic jet motion close to the line of sight, small changes in 
the intrinsic jet direction will be strongly amplified by projection effects. For example, an observed 
right-angle bend could correspond to an intrinsic bend of only a few degrees \citep[e.g.,][]{cohen2007}.
In the case of 1351-018, the jet appears to bend through $\sim130^{\circ}$ (\fref{Figure:I+p}) from a 
South-Easterly inner jet direction to a North-Westerly extended jet direction \citep[see also][]{frey2002}. 
The polarization distribution in the core region also supports an South-Easterly inner jet direction. 
This may be due to a helical jet motion along the line-of-sight as proposed, for example, for the jet 
of 1156+295 by \citet{hong2004} or small intrinsic bends due to shocks or interactions with the 
external medium amplified by projection effects in a highly beamed relativistic jet \citep[e.g.,][]{homan2002}. 
The jet bend at $\sim400$ pc from the core of 2215+020 at 5~GHz may be due to a 
shock that also causes particle reacceleration, increasing the observed jet emission.

The core degree of polarization found in these sources is generally in the range 1--3\%. 
Given that the emitted frequencies are in the range 20--45~GHz, we can compare these 
values to the core degrees of polarization observed in low-redshift AGN jets at 22 and 
43~GHz \citep{osullivangabuzda2009}. In general, the core degree of polarization is higher in the low-redshift 
objects with values typically in the range 3--7\%. This could be due to intrinsically more ordered jet magnetic 
field structures at lower redshifts or more likely, selection effects, since the 
AGN in \cite{osullivangabuzda2009} were selected on the basis of known rich polarization structure while the 
high-redshift quasars had unknown polarization properties on VLBI scales. Another likely possibility is that 
the effect of beam depolarization (different polarized regions adding incoherently within the 
observing beam) is significantly reducing the observed polarization at 5 and 8.4~GHz due to 
our much larger observing beam compared with the VLBA at 22 and 43~GHz (factor of $\sim$3 better resolution 
even considering the longer baselines in the global VLBI observations). 

Due to the extremely high redshifts, an observed RM of 50 rad m$^{-2}$ at $z\sim3.5$ (average redshift for 
the sources in our sample) is equivalent to a nearby source with an RM of 1000 rad m$^{-2}$. 
The mean core $|$RM$_{\rm min}^{\rm intr}|$ for our sample is $5580\pm3390$~rad m$^{-2}$. Naively comparing our 
results with the 8--15~GHz sample of 40 AGN reported in \cite{zt2003,zt2004}, which has a median 
intrinsic core RM of $\sim400$ rad m$^{-2}$ (and a median redshift of 0.7), suggests that the VLBI core 
RMs are higher in the early universe. However, RMs of the order of $10^4$ rad m$^{-2}$ have been 
measured in low to medium redshift sources at similar emitted frequencies 
\citep[e.g.,][]{attridge2005, algaba2010}; while some of the largest jet RM estimates, of order 
$10^5$ rad m$^{-2}$, have come through observation from 43--300 GHz of sources with 
redshifts ranging from $\sim0.1$--2 \citep{jorstad2007}. Furthermore, 15--43 GHz RM 
measurements of BL Lac objects from \cite{optical2006} and \cite{osullivangabuzda2009} 
have a median intrinsic RM of $\sim3000$ rad m$^{-2}$ and a median redshift of 0.34. 
Hence, from the small sample of minimum RM values presented in this paper it is not clear 
whether these sources are located in intrinsically denser environments or if the 
large RMs are simply due to the fact that the higher emitted frequencies are probing further 
upstream in the jet where it is expected that the electron density is higher and/or the magnetic field is stronger. 

Our results from 5 to 8.4~GHz generally show flat to slightly steep core spectral indices with a 
median value of $\alpha=-0.3$. However, there is a large range of values going from $0.7$ for 0642+449 
to $-1.0$ in the case of 1614+051. In the majority of cases, we find steep jet spectral indices ranging 
from $\alpha\sim-0.5$ to $-1.5$ with a median value of $\alpha=-1.0$. 
High-redshift objects are often searched for through their steep spectral index due to the relatively 
higher emitted frequencies for a particular observing frequency. 
The $z$--$\alpha$ correlation, as it is known, (i.e., steeper $\alpha$ at higher $z$) has been very 
successful \citep[e.g.,][and references therein]{broderick2007} in finding high-redshift radio galaxies. 
While steeper spectral index values are expected simply from the higher rest frame emitted 
frequencies, \cite{klamer2006} show that this cannot completely explain the correlation. 
The possible physical explanation they present is that the sources with the steepest 
spectral index values are located in dense environments where the radio source is pressure confined 
and hence, loses its energy more slowly. This effect might also apply to the dense nuclear 
environments of our sample of quasar jets. To test this we have plotted the core spectral index 
versus $|$RM$_{\rm min}^{\rm intr}|$ in \fref{Figure:RM_v_alpha}. The data suggest that this is a 
useful avenue of investigation with a Spearman Rank test giving a correlation coefficient of -0.95, 
but clearly more data points are needed to determine whether there truly is a correlation or not. 

Further observations of these sources with much better frequency coverage is required to properly 
constrain the spectral and RM distributions and to make detailed comparisons with low redshift 
sources to further investigate the quasar environment through cosmic time.  


\section{Conclusion}
We have successfully observed and imaged 10 high-redshift quasars in full polarization at both 
5 and 8.4~GHz using global VLBI. Model fitting two-dimensional Gaussian components to the 
total intensity data enabled estimation of the brightness temperature in the core and out along the jet. 
The observed high core brightness temperatures are consistent with modestly Doppler-boosted emission 
from a relativistic jet orientated relatively close to our line-of-sight. This can also explain the dramatic jet bends 
observed for some of our sources. 

The core degrees of polarization are somewhat lower than observed from nearby AGN jets at 
similar emitted frequencies. However, beam depolarization is likely to have a stronger effect on these 
sources compared to the higher resolution observations of nearby sources. Model-fitting the polarization 
data enabled estimation of the minimum RM for each component since obtaining the exact value of 
the RM requires observations at three or more frequencies. For sources in which we were able to 
remove the integrated (foreground) RM, we calculated the minimum intrinsic RM and found a mean 
core $|$RM$_{\rm min}^{\rm intr}|$ of 5580~rad m$^{-2}$ for four quasars with values ranging from $-1249$ 
rad m$^{-2}$ to 9110 rad m$^{-2}$. We found relatively steep core and jet spectral index values with a 
median core spectral index of $-0.3$ and a median jet spectral index of $-1.0$. 
We note that the expectation of denser environments at higher redshifts leading to larger 
RMs can also lead to steeper spectral indices through strong pressure gradients \citep{klamer2006} and that 
this hypothesis is not inconsistent with our results.  

Several of the sources presented in this paper are interesting candidates for follow-up multi-frequency observations 
to obtain more accurate spectral and RM information in order to make more conclusive statements about the 
environment of quasar jets in the early universe and to determine whether or not it is significantly different to 
similar lower redshift objects. 

\section*{Acknowledgements}
Funding for part of this research was provided by the Irish Research Council for Science, Engineering and Technology. The VLBA is a facility of the NRAO, operated by Associated Universities Inc. under cooperative agreement with the NSF. The EVN is a joint facility of European, Chinese, South African and other radio 
astronomy institutes funded by their national research councils. The Westerbork Synthesis Radio Telescope is operated by ASTRON (Netherlands Institute for Radio Astronomy) with support from the Netherlands Foundation for Scientific Research (NWO). This research has made use of NASA's Astrophysics Data System Service. 
The authors would like to thank the anonymous referee for valuable comments that substantially improved this paper. 

\bibliographystyle{mn2e}
\bibliography{highz}
\bsp

\begin{figure*}    
\begin{center} \includegraphics[width=0.7\textwidth] {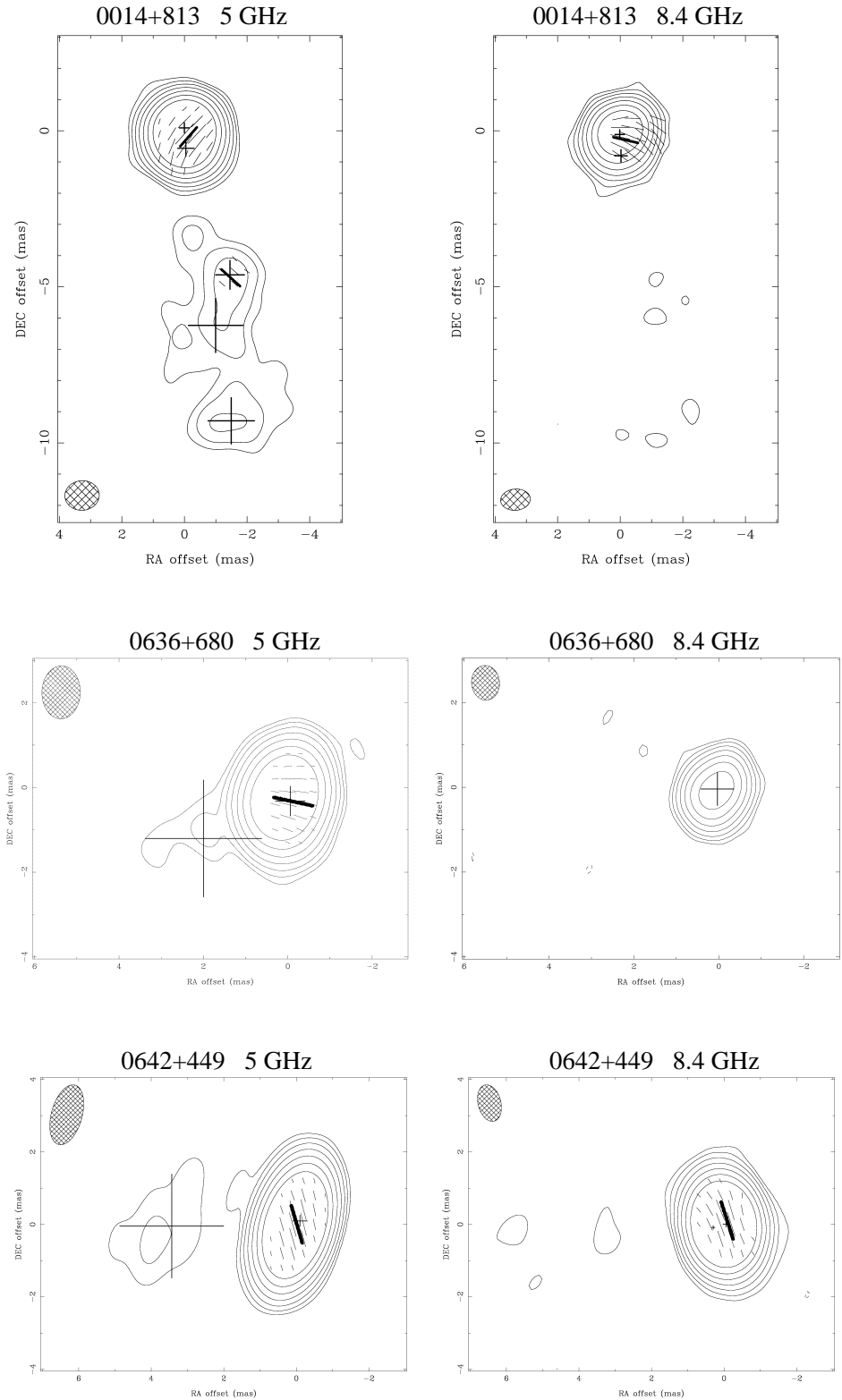}
\caption{
Total intensity ($I$) and polarization ($p$) maps for all sources at 5~GHz (left) and 8.4~GHz (right).
The image axes show the relative offset from the image centre in milliarcseconds. 
The beamsize is indicated by an inset in each image (see Table 1 for details). 
The model-fitted total intensity components are indicated by the "+" symbols while the position and angle 
of the polarized component model-fits are indicated by the boldface line. 
In all cases the contour levels increase in factors of 2 from the indicated total intensity cutoff (see Table~5).
The length of the $p$ sticks represent the relative polarized intensity. 
}
\label{Figure:I+p}
\end{center}
\end{figure*}    

\clearpage
\setcounter{figure}{1}

\begin{figure*}  
\begin{center} \includegraphics[width=0.75\textwidth] {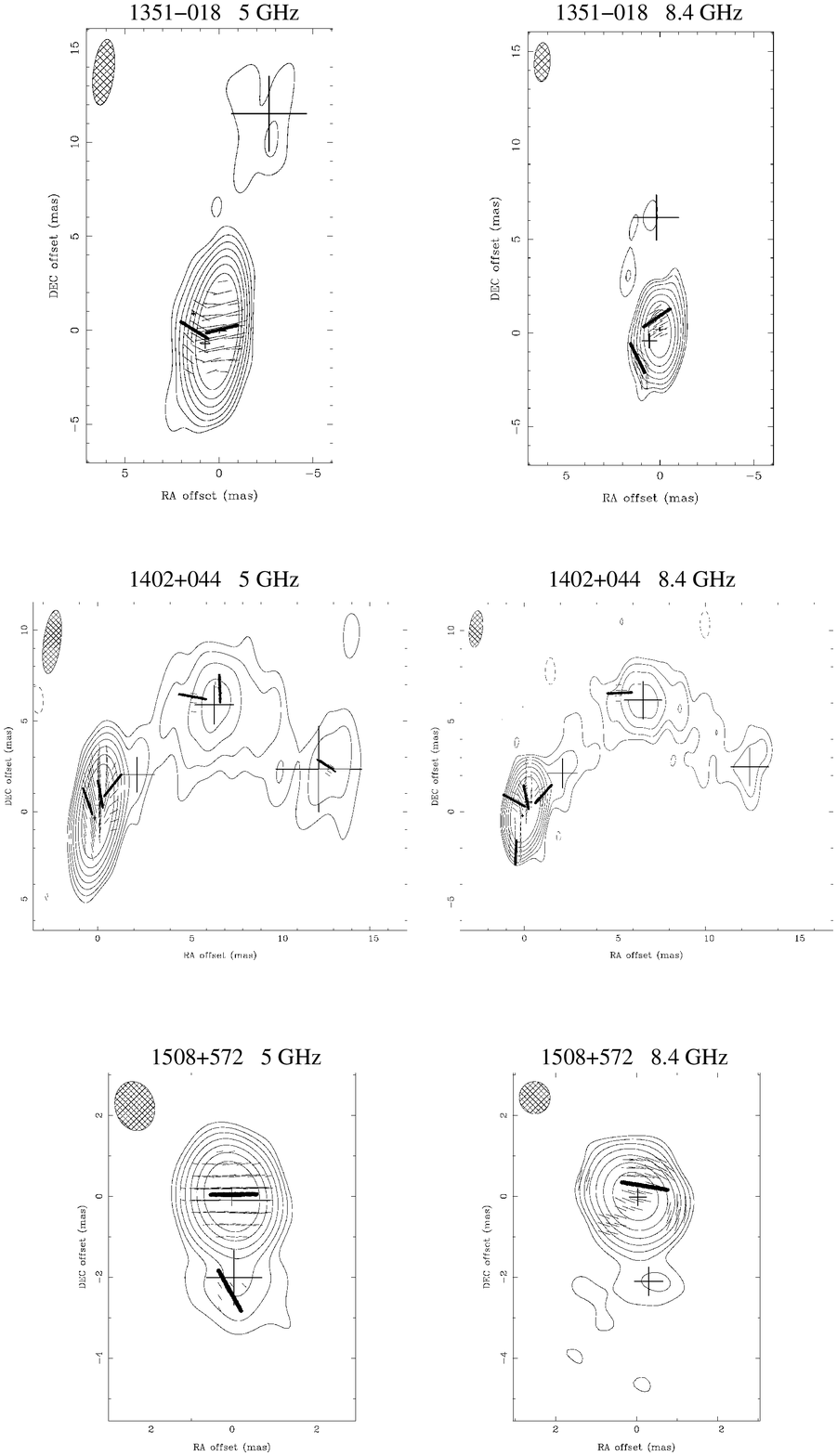}
\caption{{\it continued.}}
\end{center}
\end{figure*}    

\clearpage
\setcounter{figure}{1}

\begin{figure*}
\begin{center} \includegraphics[width=0.75\textwidth] {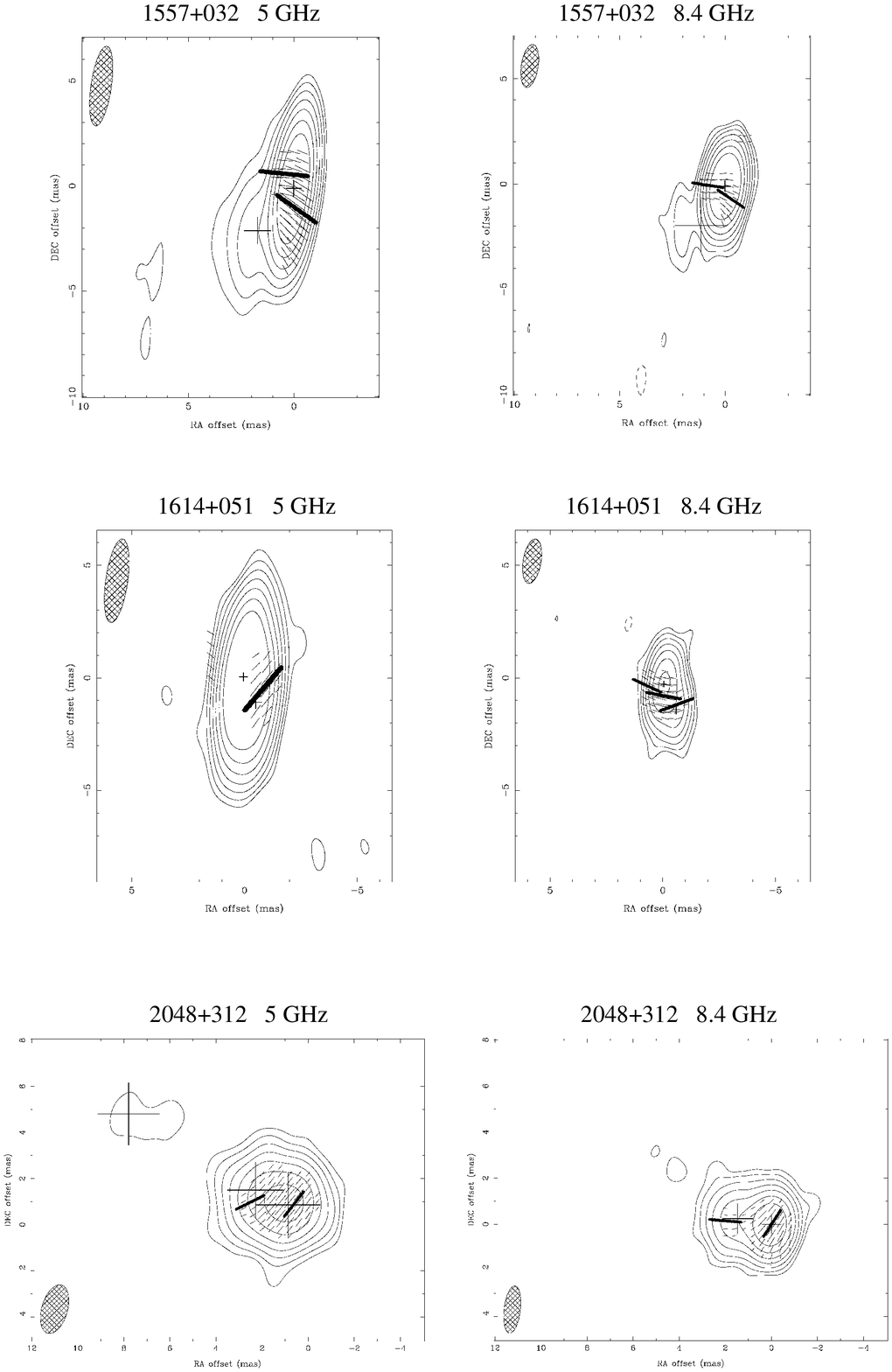}
\caption{{\it continued.}}
\end{center}
\end{figure*}    

\clearpage
\setcounter{figure}{1}

\begin{figure*}
\begin{center} \includegraphics[width=1.0\textwidth] {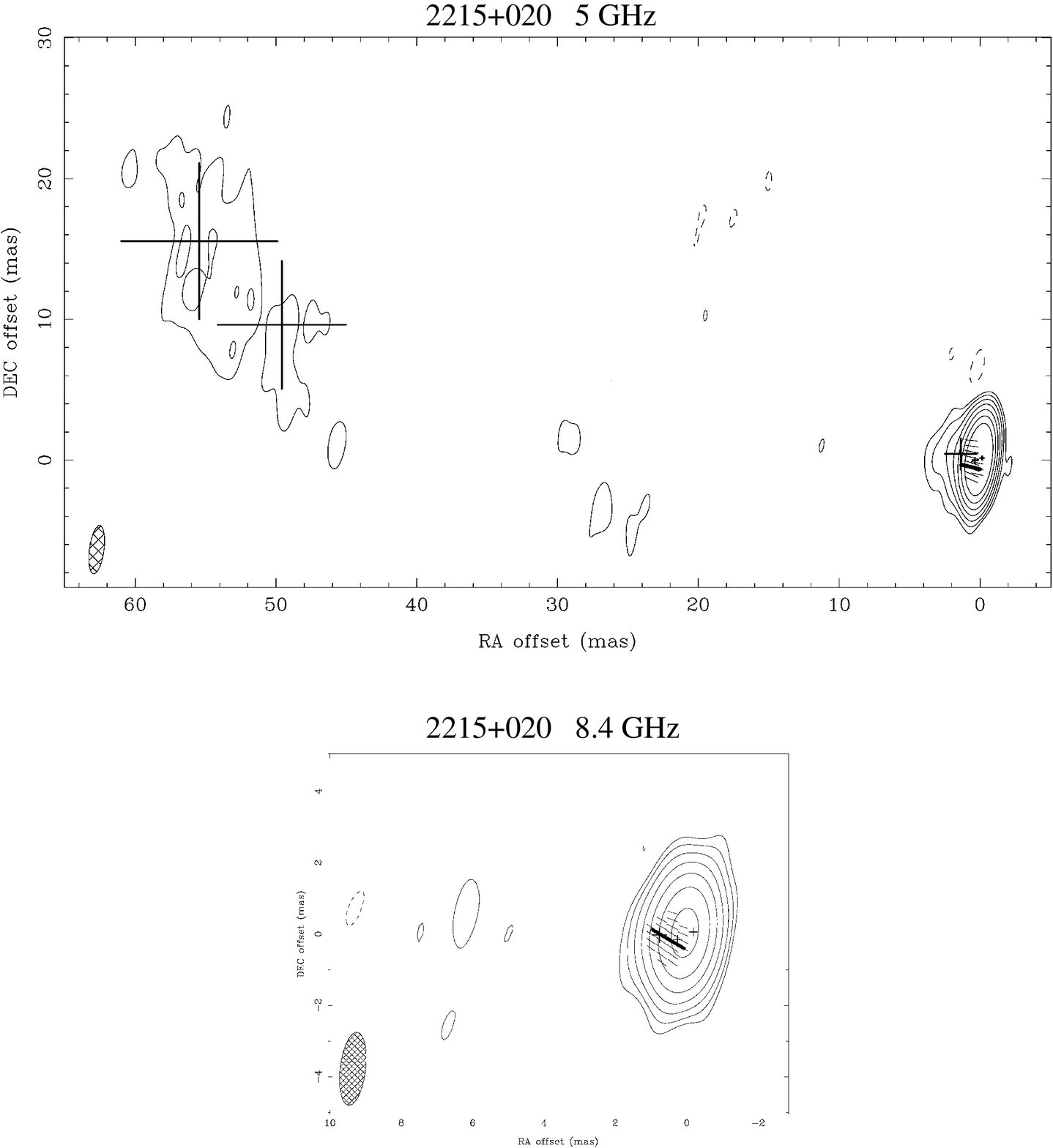}
\caption{{\it continued.}}
\end{center}
\end{figure*}

\clearpage

\begin{figure}    
\begin{center} \includegraphics[width=8cm] {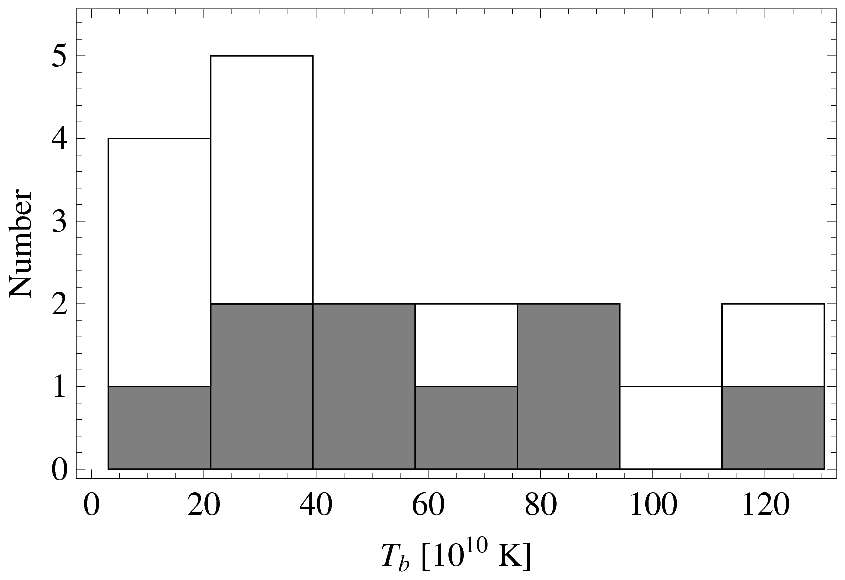}
\caption{
Histogram of core brightness temperature, in units of $10^{10} K$ at 5~GHz (grey) and 
8.4~GHz (white) for 9 sources (0642+449 excluded). 
}
\label{fig:Tb_histogram}
\end{center}
\end{figure}

\begin{figure}
  \includegraphics[width=8cm]{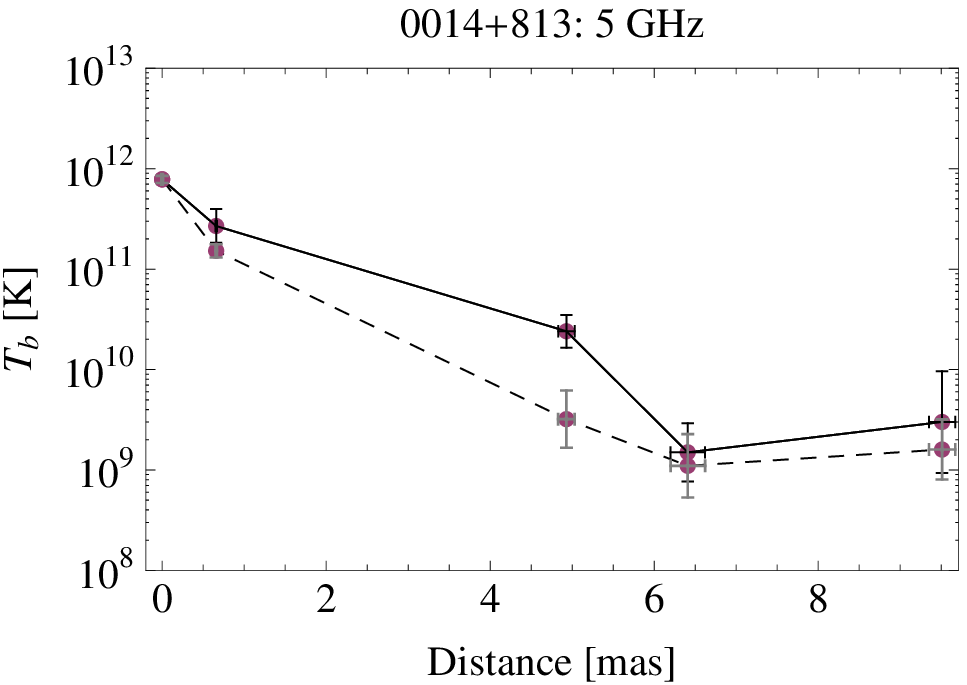}
  \caption[Brightness temperature ($T_b$) for components along the jet of 0014+813 at 5~GHz.]{Brightness temperature ($T_b$) for components along the jet of 0014+813 at 5~GHz. 
  The points connected by the dashed line represent the observed brightness temperatures 
  while the points connected by the solid line represent the model values.}
  \label{Figure:0014_Tb}
\end{figure}

\begin{figure}
  \subfigure[5~GHz data. Dashed line: observed; Solid line: model prediction with~$\alpha=-0.5$.]{
    \includegraphics[width=8cm]{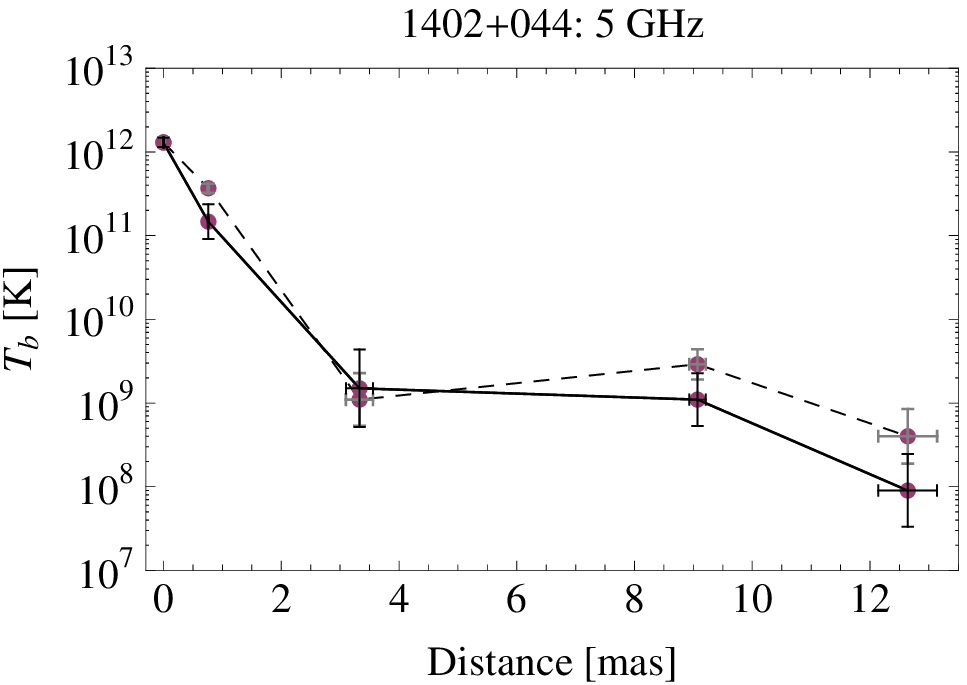}
    \label{Figure:1402_Tb_5GHz}
  }
  \subfigure[8.4~GHz data. Dashed line: observed; Solid line: model prediction with~$\alpha=-0.5$.]{
    \includegraphics[width=8cm]{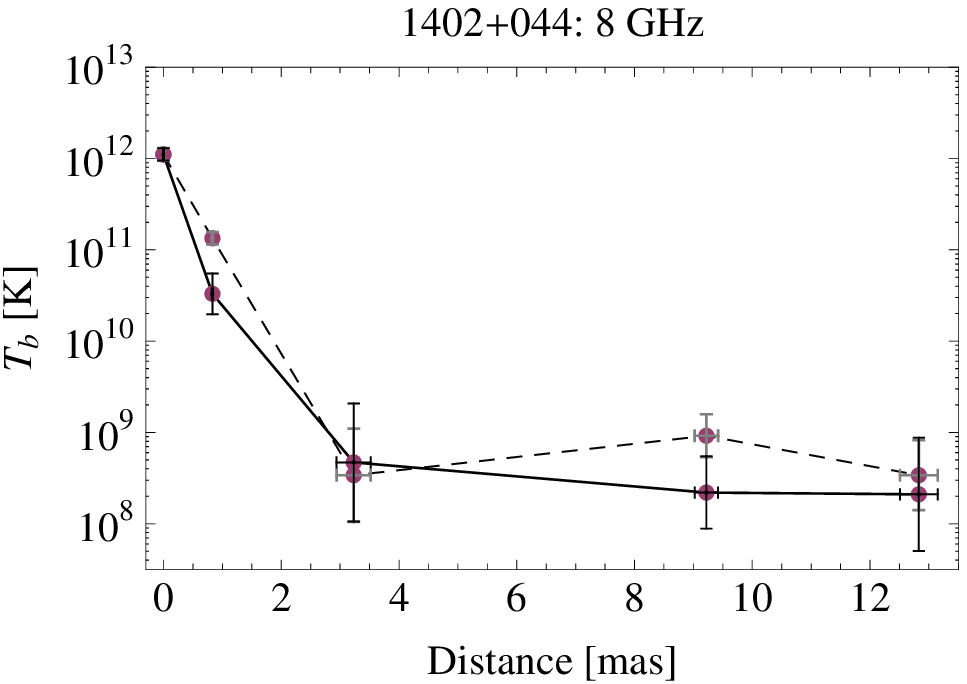}
    \label{Figure:1402_Tb_8GHz}
  }
  \caption[Brightness temperature ($T_b$) for components along the jet of 1402+044 for 
  both 5 and 8.4~GHz.]{Brightness temperature ($T_b$) for components along the jet of 1402+044 for 
  both 5 and 8.4~GHz. The points 
  connected by the dashed line represent the observed brightness temperatures 
  while the points connected by the solid line represent the model values.}
  \label{Figure:1402_Tb}
\end{figure}

\begin{figure}
  \includegraphics[width=8cm]{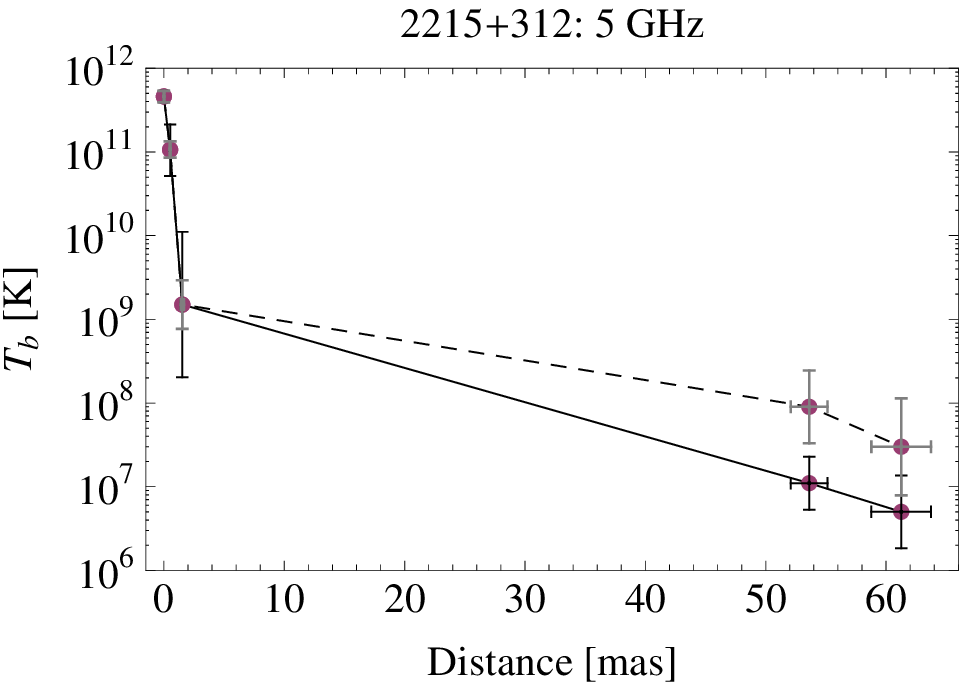}
  \caption[Brightness temperature ($T_b$) for components along the jet of 2215+312 at 5~GHz. ]{Brightness temperature ($T_b$) for components along the jet of 2215+312 at 5~GHz. 
  The points connected by the dashed line represent the observed brightness temperatures 
  while the points connected by the solid line represent the model values.}
  \label{Figure:2215_Tb}
\end{figure}

\begin{figure}
  \includegraphics[width=8cm]{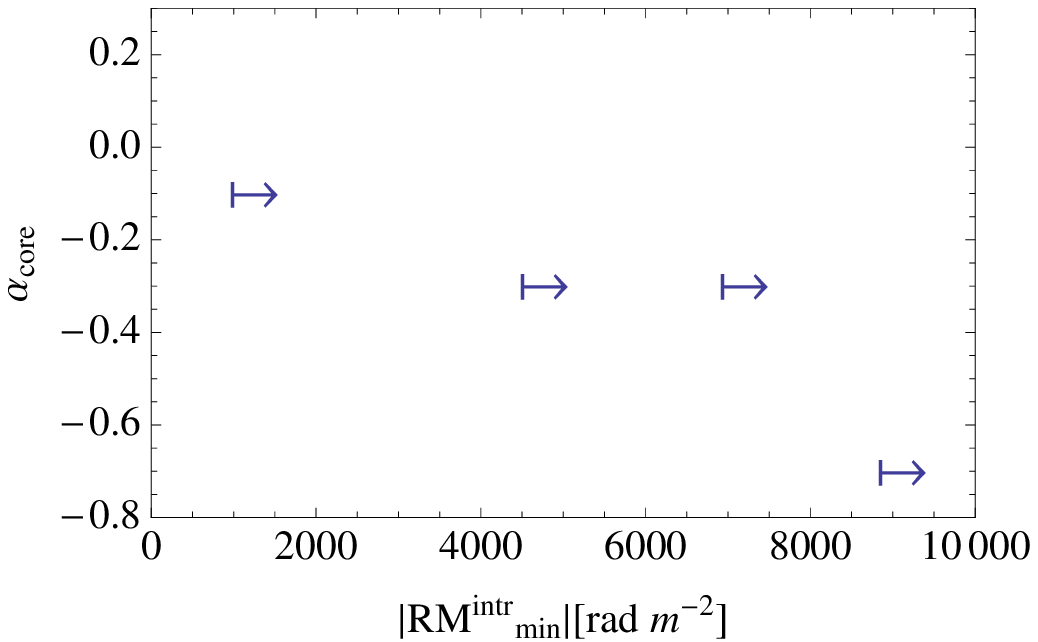}
  \caption[]{Plot of the intrinsic minimum RM (RM$_{\rm min}^{\rm intr}$) versus the core spectral index between 5 and 8.4~GHz 
  for each source. A Spearman Rank test gives a correlation coefficient of -0.95. However, more data points are required to 
  determine whether there is a correlation or not.}
  \label{Figure:RM_v_alpha}
\end{figure}

\end{document}